%% file: main.tex
\documentclass[draftclsnofoot]{IEEEtran}
\IEEEoverridecommandlockouts
\usepackage{cite}
\usepackage{amsmath,amssymb,amsfonts}
\usepackage{algorithmic}
\usepackage{graphicx}
\usepackage{textcomp}
\usepackage{xcolor}
\def\BibTeX{{\rm B\kern-.05em{\sc i\kern-.025em b}\kern-.08em
    T\kern-.1667em\lower.7ex\hbox{E}\kern-.125emX}}

\usepackage[hyphens]{url}

\usepackage{enumitem}
\usepackage{tikz}            
\usepackage{pbox}
\usepackage{pifont}
\usepackage[normalem]{ulem}

\usepackage[numbers,sort&compress]{natbib}
\usepackage[colorlinks,citecolor=red,urlcolor=blue,bookmarks=false,hypertexnames=true]{hyperref} 

\renewcommand\_{\textunderscore\allowbreak} %

\newcommand*\rectangled[1]{%
   \tikz[baseline=(R.base)]\node[draw,rectangle,inner sep=1.5pt](R) {#1};\!
}
\newcommand*\srectangled[1]{%
   \tikz[baseline=(R.base)]\node[draw,rectangle,inner sep=0.8pt](R) {#1};\!
}

\usepackage{caption}
\usepackage{subcaption}

\usepackage{comment}

\begin{document}

\title{STBPU: A Reasonably Secure \\Branch Prediction Unit}

\author{\IEEEauthorblockN{Tao Zhang}
\IEEEauthorblockA{\textit{William \& Mary} \\
tzhang06@wm.edu}
\and
\IEEEauthorblockN{Timothy Lesch}
\IEEEauthorblockA{\textit{William \& Mary} \\
tjlesch@wm.edu}
\and
\IEEEauthorblockN{Kenneth Koltermann}
\IEEEauthorblockA{\textit{William \& Mary} \\
khkoltermann@wm.edu}
\and
\IEEEauthorblockN{Dmitry Evtyushkin}
\IEEEauthorblockA{\textit{William \& Mary} \\
devtyushkin@wm.edu}
}

\maketitle
\pagestyle{plain} %

\input{abstract}
\input{intro}

\input{background}

\input{threat}

\input{design}

\input{development}

\input{analysis}

\input{evaluation}

\input{related}

\input{conclusion}

\input{acknowledgments}

\bibliographystyle{IEEEtranS}
\bibliography{references_clean}

\end{document}

%% file: abstract.tex
%% revised

\begin{abstract}

Modern processors have suffered a deluge of threats exploiting branch instruction collisions inside the branch prediction unit (BPU), from eavesdropping on secret-related branch operations to triggering malicious speculative executions. Protecting branch predictors tends to be challenging from both security and performance perspectives. For example, partitioning or flushing BPU can stop certain collision-based exploits but only to a limited extent. Meanwhile, such mitigations negatively affect branch prediction accuracy and further CPU performance. This paper proposes Secret Token Branch Prediction Unit (STBPU), a secure BPU design to defend against collision-based transient execution attacks and BPU side channels while incurring minimal performance overhead. STBPU resolves the challenges above by customizing data representation inside BPU for each software entity requiring isolation. In addition, to prevent an attacker from using brute force techniques to trigger malicious branch instruction collisions, STBPU actively monitors the prediction-related events and preemptively changes BPU data representation.

\end{abstract}

%% file: intro.tex
%% revised intro
\section{Introduction}

Although hardware attacks such as microarchitectural side channels~\cite{bernstein2005cache,percival2005cache,osvik2006cache,aciiccmez2010new,inci2016cache, aciiccmez2007predicting, ge2018survey, liu2015last}, covert channels~\cite{evtyushkin2016understanding, hunger2015understanding, naghibijouybari2016covert, maurice2015c5}, and power analysis~\cite{kocher1999differential,messerges2002examining,mangard2002simple,aciiccmez2007power,ors2004power} attacks have been known for a long time, only recently did researchers demonstrate the true power of microarchitectural attacks with newly discovered transient execution attacks, such as Meltdown~\cite{lipp2018meltdown, trippel2018meltdownprime} and Spectre~\cite{kocher2018spectre,chen2018sgxpectre,maisuradze2018speculose,kiriansky2018speculative,koruyeh2018spectreRSB,schwarz2018netspectre}. 
These attacks are based on speculative (transient) execution, a performance optimization technique present in nearly all of today's processors. While this technique improves CPU performance, with a carefully crafted exploit, it completely undermines memory protection, giving unauthorized users the ability to read arbitrary memory~\cite{kocher2018spectre,lipp2018meltdown}, bypass crucial protections~\cite{kiriansky2018speculative} or even perform arbitrary computations~\cite{evtyushkin2021computing}.

Microarchitectural attacks are possible because performance optimizations such as caches, prefetchers, and various predictors were not traditionally designed with security in mind. For example, data structures used to implement these mechanisms are often shared, making various conflicts possible. Some of these conflicts result in the leakage of sensitive data.
One such mechanism is the branch prediction unit (BPU). To maximize BPU's utilization, it
is typically shared between hardware threads; it is not flushed on mode and context switches while addresses are truncated, making it prone to various branch collisions~\cite{evers1996using, ramsay2003exploring}.
This enables attacks such as side channels~\cite{aciiccmez2007power,evtyushkin2018branchscope, evtyushkin2016jump} that are capable of leaking encryption keys or bypassing address space layout randomization (ASLR), and the recently introduced speculative execution attacks~\cite{kocher2018spectre,kiriansky2018speculative}. 
At the same time, shared BPUs are beneficial for performance. They allow high utilization of hardware structures to reduce the cost and enable efficient branch history accumulation~\cite{michaud1997trading}. 
Therefore, na\"ive protections which disable sharing or flushing BPU structures have high performance overhead.
Recently Intel introduced microcode updates implementing
countermeasures against Spectre attacks~\cite{intel-microcode}. While being effective at mitigating attacks, they can impose the performance overhead as high as 440\%~\cite{simakov2018effect, prout2018measuring}.

Despite significant efforts directed towards
designing 
various secure microarchitectural components e.g., caches~\cite{Wang:2007:NCD:1250662.1250723,kim2012stealthmem,wang2008novel,raj2009resource,coppens2009practical,varadarajan2014scheduler,liu2016catalyst} and memory buses~\cite{saltaformaggio2013busmonitor,aga2017invisimem,lehman2016poisonivy}, secure BPU designs remain
a handful of attempts~\cite{BSUP,grayson2020isca-exynos,vougioukas2019brb,zhao2021lightweight}. 
When designing a microarchitecture security mechanism, it is important to correctly estimate the attacker's capabilities.
Otherwise, it risks to be defeated by more advanced attack algorithms as was recently demonstrated with randomized caches~\cite{qureshi2019skewedceaser,purnal21PPP,bourgeat2020casa,bodduna2020brutus,song2021fixcache}.

In this paper, we propose Secret-Token Branch Prediction Unit (STBPU), a secure BPU design aimed to protect against collision-based BPU attacks and eliminate BPU side channels. 
STBPU prevents attacks by disallowing software entities from creating controlled branch instruction collisions and thus affecting
each other in an unsafe way. 
This is done by customizing the branch instruction representation for each software entity in the form of address mappings and by encrypting data stored in BPU. 
In STBPU, each software entity is provided with a unique,
randomly-generated secret token (ST) that customizes the data 
representations.  STBPU detects active attacks by monitoring related hardware events and automatically re-randomizes the ST to prevent attackers from reverse-engineering the ST value and creating collisions.

%% file: background.tex
%% merged a few subsection and re-organized to be more concise
%% removed redundant figure
\section{Background}
\label{background:main}

\subsection{BPU Baseline Model}
\label{background:org}

A typical ISA permits the following types of branch instructions: i) Direct jumps/calls where
target addresses are encoded as an offset from the current instruction pointer
and stored as an immediate value. 
ii) Conditional jumps that are only taken if a certain flag in the flag register is set.
The target of this branch is encoded similarly to direct jumps. 
iii) Indirect jumps/calls 
where targets are stored in a register or in memory, and can change throughout
program execution. iv) Return instructions are
a special type of indirect jumps where the target is stored on top of the call stack.

Below we describe a BPU baseline model that we will utilize as a foundation to build STBPU.
The baseline reflects the branch predictor (including structure sizes) used in Intel Skylake microarchitecture.  Derived from recent reverse-engineering works~\cite{evtyushkin2016jump,kocher2018spectre,zhang2020exploring, evtyushkin2018branchscope,ret2spec,koruyeh2018spectreRSB}, it represents a generalization of mechanisms used in modern Intel processors. STBPU can be applied to other branch predictor configurations and designs. This is possible because STBPU does not interfere with underlying prediction mechanisms and only changes the branch instruction representation inside BPU data structures. We demonstrate this by adapting the STBPU to protect several advanced predictors such as TAGE-SC-L~\cite{tagesclAgain} and Perceptron~\cite{perceptron}. %

\begin{figure}[]{}
     \centering
     \includegraphics[width=0.75\linewidth]{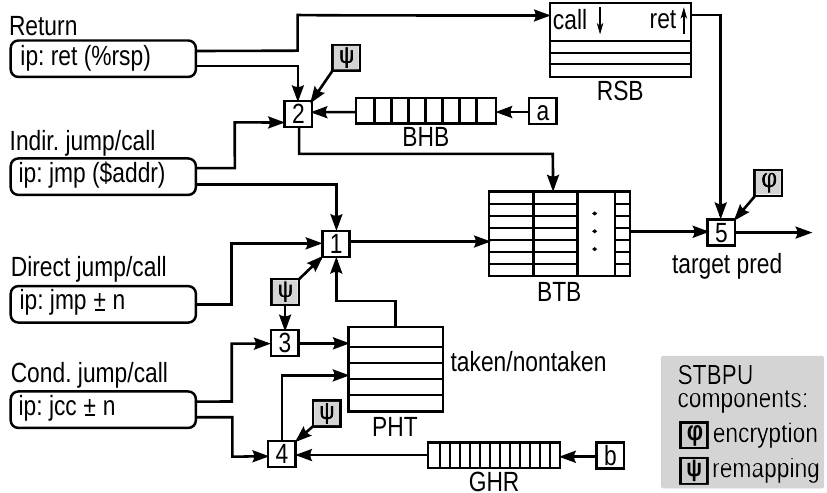}
     \caption{BPU with STBPU components highlighted}
     \label{fig:bpu}
\end{figure}

The BPU consists of the following main structures:
shift registers such as the global history register (GHR) and branch history buffer (BHB), branch target buffer (BTB), pattern history table (PHT),
and return stack buffer (RSB). 
Figure~\ref{fig:bpu} depicts how these structures are utilized during a BPU lookup with highlighted components that are modified by STBPU. The figure also shows several important functions which are referenced later. 

\noindent\textbf{Shift registers}
such as GHR and BHB are used in the BPU as a low-cost way of retaining complex branch history. GHR stores the global history of taken/not-taken branches and is used in the prediction
of conditional branches.
BHB is used by the indirect branch predictor. Its purpose is to accumulate the branch context. When a direct branch (or a call)
is executed, its virtual address is folded using XOR and mixed with the current state of BHB~\cite{kocher2018spectre}.
This context is used as part of BTB lookup, enabling BPU to predict the target of an indirect branch when it depends on the sequence of previously executed branches.

\noindent\textbf{BTB} serves the purpose of caching target addresses
of branch instructions. 
It is implemented as an 8-way, 4096-entry table. 
Each entry stores a truncated address of the 32 least significant address bits of the target.
Function~\rectangled{5} is then utilized to convert a 32-bit entry into a 48-bit 
virtual address during prediction by combining 16 higher bits
from the branch instruction pointer with 32 lower bits from BTB.
While the BTB is used to store targets for all branch types,
it has two addressing modes. In mode one, the virtual address of a branch instruction
is used to compute an index and tag. In mode two, in addition to virtual
address, the BHB is used to
perform a lookup. Mode two is only used when predicting indirect
branches, and serves as a fall-back mechanism for predicting returns.
This addressing enables storing multiple targets for a single
indirect branch depending on the context~\cite{evtyushkin2016jump,zhang2020exploring, kocher2018spectre}.

\noindent\textbf{PHT} is a large (16k entry) table consisting of 
n-bit (e.g. 2-bit) saturating counters; each counter implements a simple 
finite-state machine with states ranging from strongly non-taken to strongly-taken.
This structure is used as a base predictor to predict the direction of 
conditional branches. Previous studies~\cite{huo2019bluethunder, bhattacharya2019branch, zhang2020exploring,evtyushkin2018branchscope} indicated the presence of a mechanism similar to gshare~\cite{yeh1991two} with two distinct modes of addressing:
i) a simple 1-level mode where the virtual memory address of a branch is used
to find a PHT entry, and ii) a more complex 2-level mode where the branch
virtual memory address is hashed with global history register (GHR), enabling the accurate prediction of complex patterns.

\noindent\textbf{RSB} is used to predict return instructions.
The RSB is implemented as a fixed size (16-entry) hardware stack~\cite{ret2spec,koruyeh2018spectreRSB}.
A call instruction pushes a return address on the RSB
and a return instruction pops it.
Similarly to the BTB, RSB stores only 32 bits of the target.
Due to limited capacity, the RSB can underflow. In this case,
returns are treated as indirect branches, and the
indirect predictor is utilized for prediction.

\noindent\textbf{Microcode BPU Protections.}
Intel has proposed a set of microcode-based protections which aim to mitigate speculative execution attacks on legacy CPUs by limiting BPU structure sharing. These protections include Indirect Branch Restricted Speculation (IBRS), Indirect Branch Prediction Barrier (IBPB), and Single Threaded Indirect Branch Prediction (STIBP)~\cite{intel-microcode}. IBRS prevents higher privilege   processes from speculating with BPU data placed by lower privilege processes. This is done by flushing BPU structures when entering the kernel. IBPB provides protection by flushing the contents of the BPU on context switches. While effectively stopping BPU interference, flushing BPU structures results in a loss of useful history, causing significant performance reduction~\cite{simakov2018effect, prout2018measuring}. 
Additionally, recent research demonstrated exploitable branch collisions within same address space~\cite{zhang2020exploring,Ren21isca-Iseedeaduops}.
Thus enforcing security only during context and mode switch is not complete. 
STIBP logically segments the BPU such that the threads on the same physical core become isolated.

\subsection{BPU Attacks}
\label{back:attacks}
\input{surface-tab}

BPU can be manipulated to enable attacks of different types. For example, %
an attacker can attempt to passively observe and recover branch instruction
patterns. This happens during side and covert channel attacks. 
On the other hand, an active
attacker can manipulate the BPU state by executing branch instructions.
Such a state triggers a malicious speculative execution causing data leakage.
Moreover, attacks range based on what BPU property they utilize.
First, there are attacks that exploit the most fundamental principle of BPU 
to make predictions based on the previous behavior of a branch. E.g., if a conditional branch
was taken 100 times in a row, it is likely to be taken the next time.
An example of such an attack is Spectre-v1~\cite{kocher2018spectre}.
Second, there are attacks that exploit branch collisions (aliasing). 
Collisions appear when two different branch instructions 
map into the same BPU entry and affect one another's behavior.
In this work, we focus only on collision-based attacks. We believe that
mitigating them is an important task on its own for a number of responses. i)
There exist a large number of well-documented collision attacks
that have truly devastating effects on security~\cite{zhang2020exploring,chen2018sgxpectre,canella2019systematic,barberis2022BranchHistoryInjection,milburn2022race}.
ii) Protecting from non-collision attacks requires different principles, 
such as delaying speculative execution~\cite{sakalis2019efficient} or limiting its observability~\cite{Micro52Jiyong-STT}.
iii) Even in systems that implement safe speculation, branch collisions can still happen, causing side channel attacks. 
Because of that, we believe protecting from collision and non-collision attacks are two orthogonal tasks.

There are two BPU features that are present in nearly all CPUs 
that make collision-based attacks possible. First, the BPU data structures are
shared among all software executed on a CPU core, enabling branch
collisions between different processes. 
Second, the BPU operates with compressed
virtual addresses. For instance, out of 48 bits of branch virtual address, only 30 are utilized. Then, these bits are further compressed~\cite{kocher2018spectre}. This allows collisions to appear within the same virtual address space, e.g., collisions between branches in kernel and user process~\cite{evtyushkin2016jump}.
The deterministic nature of the BPU makes it possible for an attacker to trigger collisions in a controlled way. Our proposed solution aims at eliminating such determinism.

We detail the entire collision-based attack surface in Table~\ref{tab:surface}. 
First, we classify
attacks by where adversarial effect takes place, either within the
attacker's code (home effect) or in the victim's code (away effect).
Secondly, we classify by the kind of the effect.
A collision in BPU structures
results in either data placed by another software entity being reused,
or such data evicted and replaced. We refer to these as reuse-based and
eviction-based attacks correspondingly.
Please note that there can be different adversarial effects
enabled by same type of collision. For instance, a collision in BTB between
two different branches can result in i) BTB-data reuse, 
ii) BTB-eviction and iii) activating malicious speculative execution.
While i) and ii) results in side channel leakage of branch-related information
iii) is used as part of speculative execution attack to reveal victim's memory
contents.
As can be seen from the table, there is a diverse range of
dangerous collision-based attacks. 
By eliminating collision-based BPU attacks STBPU 
can stop many practical exploits and substantially improve security of microprocessors.

%% file: surface-tab.tex
\begin{table*}
\scriptsize
\begin{tabular}{p{3mm}|l|l|l|l|}
\cline{2-5}
& \multicolumn{2}{l|}{Reuse-based (RB)}                                                                                                                                                           & \multicolumn{2}{l|}{Eviction-based (EB)}                                                                                                                                                        \\ \cline{2-5} 
& Home effect (HE)  & Away effect(AE)  
& Home effect (HE)
& Away effect (AE)
\\ \hline
\multicolumn{1}{|l|}{Attack steps}                                                  

& \begin{tabular}[c]{@{}l@{}}
BTB:\\
1.\enskip\textbf{V}: \texttt{jmp} $s\rightarrow 	d$; BTB $\leftarrow (s,d)$ \\
2.\enskip\textbf{A}: \texttt{jmp} $s\rightarrow d'; (s,d)$ reused\\
3.\enskip\textbf{A} sees misprediction
\vspace{3pt}
\\PHT:\\
1.\enskip\textbf{V}: \texttt{jt} $s\rightarrow 	d$; PHT $\leftarrow (s,t)$ \\
2.\enskip\textbf{A}: \texttt{jnt} $s\rightarrow s+1; (s,t)$ reused\\
3.\enskip\textbf{A} sees misprediction
\vspace{3pt}
\\RSB:\\
1.\enskip\textbf{V}: \texttt{call} $s\rightarrow 	d$; RSB $\leftarrow (s+1)$ \\
2.\enskip\textbf{A}: \texttt{ret} $\rightarrow s'; (s+1)$ reused\\
3.\enskip\textbf{A} sees misprediction

\end{tabular} & \begin{tabular}[c]{@{}l@{}}
BTB:\\
1.\enskip\textbf{A}: \texttt{jmp} s$\rightarrow$d\\ 
2.\enskip\textbf{V}: \texttt{jmp} s$\rightarrow$d'\\ 
3.\enskip\textbf{V} speculatively executes $d$
\vspace{3pt}
\\PHT:\\
1.\enskip\textbf{A}: \texttt{jnt} $s\rightarrow 	d$; PHT $\leftarrow (s,t)$ \\
2.\enskip\textbf{V}: \texttt{jt} $s\rightarrow d; (s,nt)$ reused\\
3.\enskip\textbf{V} speculatively executes $s+1$
\vspace{3pt}
\\RSB:\\
1.\enskip\textbf{A}: \texttt{call} $s\rightarrow 	d$; RSB $\leftarrow (s+1)$ \\
2.\enskip\textbf{V}: \texttt{ret} $\rightarrow s'; (s+1)$ reused\\
3.\enskip\textbf{V} speculatively executes $s+1$

\end{tabular} & \begin{tabular}[c]{@{}l@{}}
BTB:\\
1.\enskip\textbf{A}: \texttt{jmp} s$\rightarrow$d; BTB $\leftarrow (s,d)$\\ 
2.\enskip\textbf{V}: \texttt{jmp} s'$\rightarrow$d'; BTB$\leftarrow(s',d')$ \\\quad$| H(s)=H(s')$,
$(s,d)$ is evicted\\ 
3.\enskip\textbf{A} sees $s$ mispredicted
\vspace{1pt}
\\PHT:\\
\textit{PHT entries are not evicted}
\vspace{3pt}
\\RSB:\\
1.\enskip\textbf{A}: \texttt{call} $s\rightarrow d$; \\RSB $\leftarrow (s+1)$ then fills RSB \\
2.\enskip\textbf{V}: \texttt{call} $s'\rightarrow 	d'$; \\RSB $\leftarrow (s'+1)$ evicting $(s+1)$\\
3.\enskip\textbf{A} sees misprediction

\end{tabular} & \begin{tabular}[c]{@{}l@{}}
BTB:\\
1.\enskip\textbf{V}: \texttt{jmp} $s \rightarrow d$; BTB $\leftarrow (s,d)$\\ 
2.\enskip\textbf{A}: \texttt{jmp} $s' \rightarrow d'$; BTB$\leftarrow (s',d')$ \\\quad$| H(s)=H(s')$\\ 
3.\enskip\textbf{V}: CPU uses static prediction
\vspace{3pt}
\\PHT:\\
\textit{PHT entries are not evicted}
\vspace{3pt}
\\RSB:\\
1.\enskip\textbf{V}: \texttt{call} $s\rightarrow d$; RSB $\leftarrow (s+1)$\\
2.\enskip\textbf{A}: overflows RSB by \\looping \texttt{call} $s'\rightarrow d'$\\
3.\enskip\textbf{V}: CPU uses static prediction

\end{tabular} 
\\ \hline
\multicolumn{1}{|l|}{\begin{tabular}[c]{@{}l@{}}Adversarial\\ effects\end{tabular}} 
& \begin{tabular}[c]{@{}l@{}}Visible source and target of branch/call\\ addresses, 
taken/nontaken\\patterns~\cite{aciiccmez2007predicting,evtyushkin2018branchscope,lee2017inferring,evtyushkin2016jump,koruyeh2018spectreRSB}\end{tabular} 
& \begin{tabular}[c]{@{}l@{}}Timing channel due to \textbf{A}\\ controlling predictions
in \textbf{V}~\cite{aciiccmez2007predicting},
\\speculative execution attacks
\\\cite{kocher2018spectre,koruyeh2018spectreRSB,chen2018sgxpectre,ret2spec,schwarz2018netspectre, zhang2020exploring}\end{tabular} & 
\begin{tabular}[c]{@{}l@{}}\textbf{V}'s \texttt{jmp} taken/nontaken~\cite{aciiccmez2007predicting} and 
\\\texttt{call} pattens, branch \\instruction virtual address~\cite{kocher2018spectre}\end{tabular}
& \begin{tabular}[c]{@{}l@{}}Timing channel due to \textbf{A} forcing\\ static default predictions~\cite{aciiccmez2007predicting}, 
\\speculatively execute gadget 
\\at static prediction address~\cite{canella2019systematic}
\end{tabular}
\\ \hline
\end{tabular}
\centering

 \textbf{A}: attacker; \textbf{V}: victim; \texttt{jmp} $s\rightarrow d$: jump from $s$ to $d$; \texttt{call} $s\rightarrow d$: call function $d$ from callsite $s$; BTB/PHT/RSB$\leftarrow (s,d)$: store target $d$ for branch $s$ in BTB/PHT/RSB; $H()$: BTB/PHT hash function; $s+1$: next instruction after $s$

\caption{Attack surface classification for BPU collision-based attacks by event and adversarial effect types}
\label{tab:surface}
\end{table*}

%% file: threat.tex
\section{Threat Model}
\label{threat:main}
We assume a powerful attacker that
has a \textit{complete} understanding of all hardware components and structures in the STBPU. The attacker has access to normal reverse engineering resources, such as time measurements and performance counters, and has access to a wide variety of hardware covert channels. The STBPU design calls for new special purpose registers as detailed in Section~\ref{design:main}; the adversary is assumed to be unable to read/modify the contents of these registers. Such a role is delegated to a privileged software entity (OS, hypervisor) which attacker does not control. 

We assume the attacker cannot gain access to ST for the victim process neither when it is in the special purpose register, nor when system software stores it. Former is impossible because the register can only be accessed from the privileged mode. Later happens only in the event of system software compromise. The ST can be considered as part of processes's context that is saved and restored on context switches. The event of attacker gaining access to context data would be equal to a full compromise. In such case, there is no point for attacker to use side channels.

\noindent
We consider attacks presented in Table~\ref{tab:surface} including both side channel attacks in which victim executes a sensitive data dependent branch branch
as well as speculative execution attacks where victim is forced to speculatively execute leakage gadget code. 
We assume the following two attack scenarios:

\noindent\textbf{Sensitive Process as Victim.}
In this scenario, an attacker tries to learn sensitive data from a victim process by manipulating the BPU state and recording observations. The attacker has control over user-level process co-located on the same CPU core and is capable of performing activities that are normally allowed to untrusted process such as accessing fine grain hardware timers via~\texttt{rdtscp} instructions. We assume the victim and attacker can either execute on two logical cores within the same physical core or share the same logical core with time-slicing. This scenario also includes recently introduced transient trojans~\cite{zhang2020exploring} where collisions occurring within the same memory segments are exploited.

\noindent\textbf{Kernel/VMM as Victim.}
The attacker takes a form of a software entity with lower privilege level, i.e. untrusted user process. The attacker tries to learn sensitive data owned by a higher privileged entity (OS kernel or VMM) by manipulating with BPU state and recording observations. Here, victim and attacker share a same contiguous virtual address space. Attacker is restricted from executing privilege instructions.

%% file: design.tex
%% revised

\section{STBPU Design}
\label{design:main}

As discussed in Section~\ref{background:main}, BPU attacks are possible
due to deterministic mapping mechanisms, allowing attackers to create branch collisions.
STBPU aims to stop these attacks by replacing these deterministic mechanisms with keyed remapping mechanisms which prevent branch collision construction. 
The design philosophy of STBPU is to create different data representations for separate software entities inside the BPU data structures.
Each software entity requiring isolation is assigned a unique
ST, which is a random integer that controls how branch virtual addresses
are mapped into BPU structures. This ST is also used to encrypt/decrypt stored data. 
Compared to na\"ive protections based on flushing or partitioning, our approach has a number of benefits.

Consider a protection scheme where branch target poisoning is prevented by
flushing the BTB on context switches. Invalidating the entire branch target
history will negatively affect performance in cases where context switches
are frequent. Similarly, to protect from target collisions between kernel and user branches, BTB must be flushed on mode switches (e.g. all syscalls).
Partitioning hardware resources reduces the effective capacity of BPU structures 
resulting in a higher miss rate and lower prediction accuracy.  
Instead, a customized mapping approach allows separate
software entities to co-exist in the BPU with minimal performance overhead; performance evaluated in Section~\ref{eval:main}. STBPU utilizes two key approaches to enable safe resource sharing.

\begin{itemize}
\item We make collision creation difficult
by ensuring all remapping functions are dependent upon both branch address and ST. 

\item STBPU detects when a potential
attacker process has recovered sufficient information
that enables deterministic collision creation by monitoring hardware events.
\end{itemize}

\subsection{ST re-randomization} 
The ST of the current process in the BPU is re-randomized once
certain (OS controlled) thresholds are reached. 
Note that in STBPU design, the OS is trusted and is responsible for setting parameters such as the re-randomization threshold. This is a common assumption for systems protecting against microarchitectural attacks since compromising OS gives the attacker full control over the system, making such attacks non-necessary. On the other hand, such a design choice makes our mechanism more flexible and permits the OS to adjust the strength of enforcement based on factors such as whether a certain process is considered sensitive or the attacker's capabilities. For instance, if a more effective side channel attack is discovered after STBPU is deployed, the underlying hardware mechanism will still remain effective and will only require the OS to readjust the thresholds. Moreover, for the extreme cases of sensitive processes the OS may opt to set the threshold as low as 1, forcing re-randomization after every branch instruction, effectively disabling the BPU mechanism.

STBPU can be also adapted for systems with OS not trusted (e.g. SGX), then another system component needs to be responsible for managing tokens and thresholds. For instance, in the case of SGX, the enclave entering routine can serve this purpose. Alternatively, simple logic of ST management in STBPU should also enable hardware only implementation. 
Re-randomizing ST effectively resets the customization of the BPU data representation for that process. 
Although it leads to the loss of branch history (by making it unusable), our analysis indicates that such events are
infrequent. Re-randomizing the ST of one process does not remove stored history of a process with a different ST. This is the key difference compared to flushing-based approaches.
We derive the re-randomization thresholds through the analysis in Section~\ref{analysis:main}.

While potentially dangerous, branch history sharing between programs
benefits performance. Consider a server application
that spawns a new process for each incoming connection.
Since each process executes the same code, the accumulated
BPU state is used by the newly spawned process. This allows the new process
to avoid the lengthy period of BPU training.
STBPU permits selective history sharing
by allowing the OS to provide multiple copies of the same program to utilize the same ST value.
However, when sharing is not desired, each thread can be given a unique ST.

\subsection{Hardware Mechanisms and Interfaces}
\label{design:STBPU}
Since current BPU designs are highly optimized in terms of performance and hardware cost,
we restrict ourselves to only modifying BPU mapping mechanisms,
adding registers, and encrypting stored targets. 
Such changes will provide similar performance to the unprotected design
and make STBPU agnostic to a particular BPU design.
In STBPU, each hardware thread is provided with an extra register to store the ST of the current process. 
Only the OS is allowed to read/modify these registers, and these registers are inaccessible in unprivileged CPU mode. 
As such, the OS facilitates history retention across context/mode switches by loading the appropriate STs. 
We also add several model-specific registers (MSRs) that store thresholds and counters for automatic
ST re-randomization. These MSRs monitor the events that indicate an active
attacker process. 
We monitor two events: i) branch mispredictions which includes incorrectly
predicted direction of conditional branches and targets of any branch, and ii) BTB evictions.
In Section~\ref{analysis:main}, we explain how these events are utilized to deter BPU attacks.
Initially, the counter values are set to their respective threshold values.
When an event is observed, the corresponding counter is decremented. 
When a counter reaches 0, the current ST is re-randomized, and the CPU reset the counter with the threshold value.
The OS treats these registers as a part of software context saving, and recovering their values
on context/mode switches. We assume re-randomization is done by fetching a value from
low-latency in-chip pseudo-random number generator~\cite{m_2019}. 

\begin{table}[t]
\scriptsize
\centering
\resizebox{\columnwidth}{!}{%
    \begin{tabular}{c|c|c|c|c|}
    \cline{2-5}
    & Baseline input  & STBPU input & Output & Function  \\ \hline
    \multicolumn{1}{|c|}{\hskip 2pt \srectangled{1}\quad} & $32$ $s$ & 32 $\psi$, 48 $s$ & 9 \texttt{ind}, 8 \texttt{tag}, 5 \texttt{offs} & $R_{1}(80\mapsto22)$ \\ \hline
    \multicolumn{1}{|c|}{\hskip 2pt \srectangled{2}\quad} & 58 \texttt{BHB} & 32 $\psi$, 58 \texttt{BHB} & 8 \texttt{tag} & $R_{2}(90\mapsto8)$ \\ \hline

    \multicolumn{1}{|c|}{\hskip 2pt \srectangled{3}\quad} & 32 $s$ & 32 $\psi$, 48 $s$ & 14 \texttt{ind} & $R_{3} (80\mapsto14)$ \\ \hline
    \multicolumn{1}{|c|}{\hskip 2pt \srectangled{4}\quad} & 18 \texttt{GHR}, 32 $s$ & 32 $\psi$, 16 \texttt{GHR}, 48 $s$ & 14 \texttt{ind} & $R_{4} (96\mapsto14)$ \\ \hline
    \multicolumn{1}{|c|}{\hskip 2pt \srectangled{t}\quad} & 48 $s$, \texttt{L(GHR)} & 32 $\psi$, 48 $s$, \texttt{L(GHR)} & 10/13 \texttt{ind}, 8/12 \texttt{tag} & $R_{t} (80{\uparrow}\mapsto25)$ \\ \hline
    \multicolumn{1}{|c|}{\hskip 2pt \srectangled{p}\quad} & 48 $s$ & 32 $\psi$, 48 $s$ & 10 \texttt{ind} & $R_{t} (80\mapsto10)$ \\ \hline
    \end{tabular}
}
\\
\texttt{L(GHR)} --- represents geometric series of global history lengths
\newline
\texttt{s} --- represents the source bits of branch instructions %

\caption{I/O bits for baseline and STBPU functions}
\label{tab:stbpu_bits}
\end{table}

The ST register is a 64-bit register divided into two 32-bit chunks, $\psi$ and $\varphi$.
The first chunk $\psi$ acts as a key for a keyed remapping functions making BPU mapping unique for each process. 
We replace functions \rectangled{1}, \rectangled{2}, \rectangled{3} and \rectangled{4}
in Figure~\ref{fig:bpu} with STBPU remappings $R_{1..4}$ accordingly.
We add functions $R_{t}$ and $R_{p}$ that are used for STBPU implementation with the TAGE and Perceptron predictors.
Both baseline and STBPU remapping functions reduce input data (address, BTB, GHR bits) into fixed size index, tag, and offsets
used by the BPU to perform lookups.
Section~\ref{development:automation} describes how $R_{1..4,t,p}$ were selected.
Additionally, these functions utilize the entire 48-bit virtual address unlike legacy functions that use truncated address bits as inputs.
This is crucial to prevent the same address space attacks~\cite{zhang2020exploring}. Table~\ref{tab:stbpu_bits} details all input/output bit changes between the baseline and STBPU models.

We use a simple scheme based on XOR to encrypt data stored in BPU structures
to stop attackers from 
redirecting execution to a desired speculative gadget even if collisions occur.
In the case of a collision, speculative execution will be redirected to an encrypted (random) address. This will effectively stall
malicious speculative execution.
In STBPU, every entry stored in BTB and RSB is XORed with $\varphi$ of the current process. Note that the baseline BPU stores only 32 bits of target addresses,
so the 32-bit $\varphi$ is sufficient for encrypting all stored bits.
We use a simple XOR encryption for two reasons: i) XOR operations are extremely fast with trivial hardware implementation, and 
ii) automated ST re-randomization makes the simple XOR encryption sufficiently strong (discussed in Section~\ref{analysis:main}).
To decrypt data
in BTB and RSB, we modify the function \rectangled{5}, which XORs target bits with $\varphi$ before extending them to 48-bit address.

%% file: development.tex
% revised
\section{Implementation}
\label{development:main}
In Section~\ref{design:main}, we defined remapping functions $R_{1..4,t,p}$ which replace the methods of calculating indexes, tags, and offsets for lookup purposes in the baseline BPU model. Remapping functions $R_{1..4,t,p}$ can be thought of as \textit{non-cryptographic} hash functions.
Given the size constraints of the BPU structures, collisions between different inputs to functions $R_{1..4,t,p}$ will occur; this fact prevents functions $R_{1..4,t,p}$ from providing cryptographic security, regardless of implementation.
This inherent weakness is remedied with periodic re-randomization of STs; the security of such re-randomizations are discussed in Section~\ref{analysis:main}.  The mapping functions used in the baseline model are not fully reverse engineered, but we can safely assume some fast compression functions are used with delays of no more than 1 clock cycle.
Using performance and security as our guides, we placed several important constraints upon functions $R_{1..4,t,p}$:

\begin{enumerate}[label=C\arabic*]
\item \label{dev:c1}The compute delay for $R_{1..4,t,p}$ must not exceed $C$ clock cycles, where $C$ may vary from CPU to CPU.  For our purposes, we choose $C$ to be 1 clock cycle. 
We enforce this by limiting the number for transistors of each remapping function on the critical path.
\item \label{dev:c2}The function must provide \textit{uniformity}: outputs of $R_{1..4,t,p}$ should be uniformly distributed across their respective output spaces.
\item \label{dev:c3}The function must demonstrate \textit{avalanche effect}~\cite{noncryptohashes}: The outputs of $R_{1..4,t,p}$ must appear to be pseudo-random, and the relationship between inputs and outputs should be non-linear.

\end{enumerate}

We analyzed existing hardware supported hashing mechanisms, but found none that satisfied our specific requirements. Specifically, existing multi-round
hash functions exceed the single CPU cycle constraint. Later we describe a mechanism we developed to automatically generate remapping functions
taking into account aforementioned constraints. In addition to remapping, STBPU requires encryption of branch addresses stored inside BPU.
We found out that existing lightweight cryptographic functions are not suitable for our purposes for two main reasons:
First, using strong ciphers does not directly translate into better security which are primarily designed to withstand known plaintext/ciphertext attacks. However, STBPU threat model is much different as attackers never observe encrypted addresses (ciphertext) nor partially matched plaintext/ciphertext. They only observe collisions (not knowing with their own or victim's branch) and need to reverse-engineer the rest of the address bits. Besides, knowing their own STs does not provide immediate access to collision creation or simplifies collision-based attacks. 
In Section~\ref{analysis:main}, we show that the number of mispredictions and evictions attackers must incur to successfully 
infer
a ST far exceeds the thresholds that will trigger ST re-randomization. Thus, encrypting with a more advanced cipher would not increase the 
level of security. Secondly, more sophisticated encryption schemes introduce significant delays in CPU frontend. For instance, we explored PRINCE-64~\cite{borghoff2012prince} and Feistel-Network~\cite{menezes2018handbook} to encrypt stored branch targets. While comparably fast, PRINCE-64 and Feistel-Network will still consume multiple clock cycles and consume more energy due to higher number of gates compared to a simple subsingle-cycle XOR operation.

\subsection{Automation of Finding Remapping Functions}
\label{development:automation}

\noindent\textbf{Automated Remap Generation Algorithm.}
Designing the remapping mechanisms is a multi-variable optimization problem. 
To solve it, we developed an algorithm that takes in a list of hardware constraints, and randomly generates remapping function candidates. 
The algorithm composes the function from a predetermined pool of primitives. Each remapping function is iteratively generated and tested one layer at a time, where a layer is a block of these primitives. After a layer is added, the current function is tested against the supplied constraints.  There are three possible scenarios that occur during each round of testing. i) The current design satisfies all constraints, and subsequently stored for later optimization. ii) The current design violates one or more constraints, and is discarded. iii) The current design does not outright violate the constraints, but is incomplete.  In case 3, our algorithm changes the weights used for primitive selection during the creation of the next layer to improve the current design.

\noindent\textbf{Constraint Selection of}~\ref{dev:c1}.
Our algorithm requires an input of several variable constraints for the generated remapping functions to satisfy~\ref{dev:c1}.  These constraints are: the maximum count of transistors along the critical path, the maximum number of transistors in parallel (breadth), the maximum number of total transistors for the design, the number of input and output pins, the maximum number of functional layers (blocks) the design can have,  and the maximum number of wires an arbitrary wire can cross over. 

Modern processors are designed to perform 15-20 gate operations in a single cycle~\cite{CEASER2018}, which translates to roughly 30-45 transistors along the critical path.  The delay incurred by each transistor in the critical path is relatively independent of the CPU clock cycle; therefore, the faster the CPU's clock cycle, the smaller the number of transistors that can be completed within 1 clock cycle.  Therefore, we assume 45 is the absolute maximum number of transistors we allow in the critical path with preference set for shorter critical paths. 

\noindent\textbf{Primitive Selection.}
Much research has been conducted into cryptographic hash primitives~\cite{Leander:2007:CBS:1420233.1420250,10.1007/978-3-642-23951-9_21,10.1007/978-3-540-74735-2_31, newSboxes, Zhang2015} that provide building blocks for hash functions with strong properties. 
We leverage these primitives from SPONGENT~\cite{10.1007/978-3-642-23951-9_21} and PRESENT~\cite{10.1007/978-3-540-74735-2_31} hashes. Out of those S-boxes (establishing non-linearity by substations) are perhaps most critical. %
To increase the simplicity of remapping function generation, we separate primitives into two categories: non-invertible compression primitives and mixing primitives.

Non-invertible primitives tend to employ XOR logic gates to obfuscate the relationship between input and output. For many such primitives, multiple inputs generate the same, smaller output which makes reverse-engineering difficult. Combining multiple non-invertible layers increases complexity of attacks aiming to pair a known output to an unknown input. These primitives \textit{compress} input size $|m|$ to an output size $|n|$ where $|m|>|n|$.  Table~\ref{tab:stbpu_bits} shows the disparity between the input and output sizes for $R_{1..4,t,p}$ functions, and indicates the need for optimized compression primitives.  Mixing primitives are primarily used to introduce \textit{non-linearity} to a hash design which makes deterministically changing the output by varying the input difficult. These primitives are primarily composed of $|m|\mapsto |m|$ sized S-boxes and P-boxes (performing permutations). Since the hardware complexity of S-boxes increases superlinearly with the size of $|m|$, we limit our S-boxes to a maximum of 4 input/outputs.  These S-boxes can be implemented efficiently with combinatorial logic or transistor/diode matrices. P-boxes are constrained by the maximum wire crossover set for the algorithm.

\noindent\textbf{Validation of \textit{Uniformity} (\ref{dev:c2}) and \textit{Avalanche Effect} (\ref{dev:c3})}

Remapping functions that satisfy the hardware constraints are then tested against constraints~\ref{dev:c2} and~\ref{dev:c3}.
We first employ the balls and bins analysis and compute the coefficient of variation (CV) of bins to approximate the uniformity (\ref{dev:c2}) of the output space~\cite{BallsAndBinsRaab:1998:BBS:646975.711521}.  
\ref{dev:c3} is satisfied when a remapping adheres to a strict avalanche criterion.
To quantify the avalanche effect of $F$, for each input $\lambda$, we generate a set of unique inputs, $S$, where each input in $S$ differs from $\lambda$ by a single bit flip.  We then compute the hamming distance between $F(\lambda)$ and $F(S_i)$, for all inputs in $S$.  Using these hamming distances, we determine the CV of the hamming distances for a particular $\lambda$. We test each $F$ with 1 million random inputs and compute the average hamming distance for all inputs.  The ideal case occurs when: i) the average hamming distance over 1 million random inputs is roughly 50\%. ii) For all inputs, the CV of the average hamming distance for each input is 0.  iii) For all bit positions of an output of $F$, the difference between the minimum and maximum hamming distances for a bit flip in any bit position is 0.

\subsection{Optimization and Remapping Selection}
\label{development:optimization}

\begin{figure}[t]{}
     \centering
     \includegraphics[width=0.85\linewidth]{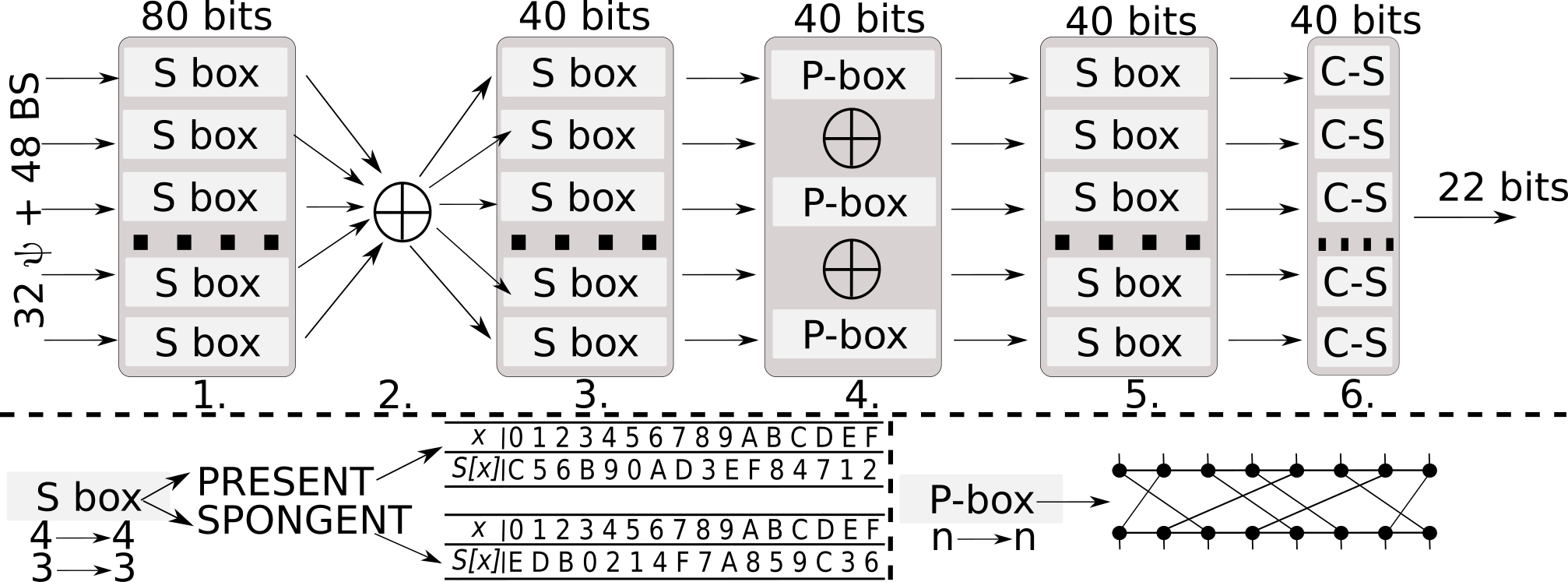}
     \caption{$R_{1}$ remapping function construction}
     \label{fig:stbtb_remap_func}
\end{figure}

The final selection of remapping functions $R_{1..4,t,p}$ is primarily based upon the results from the previous tests.  The result is a multiobjective optimization problem where the ideal state for different desired metrics may be maximized or minimized.  To make all metrics comparable, we normalized each metric so that the optimal value is 0.  
We then considered this to be a simple weighted optimization problem where we seek functions that yield the lowest sum of all metrics recorded when testing for uniformity and the avalanche effect.  Let $F$ be a particular function in the group of potential functions $G$ for remapping function $R_{i}$, for $i \in R_{1..4,t,p}$:

\begin{equation}
min\sum_{i}^{k}w_ig(F), F\in G \label{eq:0}
\end{equation}
All weights were set to 1 to avoid prioritizing one metric over another. Further prioritizing then can be done by hardware developers for a specific CPU design. 
For space reasons, we do not show the designs for all of $R_{1..4,t,p}$ since they share many similar characteristics.  Instead, we show the chosen design for $R_{1}$ in Figure~\ref{fig:stbtb_remap_func} where stages 1, 3, and 5 are substitution layers using $4 \mapsto 4$ and $3 \mapsto 3$ S-boxes.
 For space reasons, not all types of S-boxes are shown.  Under the design of $R_{1}$, we show the logical mappings for S-boxes used by PRESENT and SPONGENT. P-boxes are $n \mapsto n$ in size with the pin mappings generated randomly by our remap function generator.
 C-S boxes are compression structures that map $|m|$ bits to an output size of $|n|$ bits where $|m|>|n|$.
 This design of $R_{1}$ has a critical path length of 36 transistors, so it is capable of being computed within a single clock cycle.

%% file: analysis.tex
%% removed redundant content and table; added new analyis to DoS type of attacks
%% explained more on why existing random desgins are not suitable in BPU defense
\section{Security Analysis}
\label{analysis:main}

We assume any attackers can have complete knowledge of all STBPU remapping functions, full control of execution flow, and are capable of executing branches to/from any address within their processes. The goal is to enable malicious branch instruction collisions that allow mounting one of the collision-based attacks. STBPU makes collisions non-deterministic, forcing the attackers to rely on either brute force approaches or reverse-engineering the ST value. Further, attackers can utilize recently proposed fast attack algorithms such as GEM~\cite{qureshi2019skewedceaser} and PPP~\cite{purnal21PPP} that target randomized caches~\cite{bodduna2020brutus,bourgeat2020casa}.

\begin{table}[h] %
\resizebox{\columnwidth}{!}{%
    \begin{tabular}{|l|l|}

    \hline
    \textbf{Parameter: Description}                             & $A$: Branch in attacker(A)'s address space     \\ \hline
    $W_{struct}$: Number of ways                    & $V$: Branch in victim(V)'s address space       \\ \hline
    $I_{struct}$: Number of sets (indexes)          & $\psi_{a/v}$: A/V  $R()$ 32-bit token          \\ \hline
    $T_{struct}$: Entry tag bit entropy             & $\varphi_{a/v}$: A/V  target encryption token  \\ \hline
    $O_{struct}$: Entry offset bit entropy          & $\tau_{Q}$: Target of arbitrary branch $Q$     \\ \hline
    $\Omega_{struct}$: Entry target bit entropy $Q$ & $E_{Q}$: Entry stored for arbitrary branch $Q$ \\ \hline
    \end{tabular}
}
\caption{Parameters used in STBPU analysis}
\label{tab:ids-st}
\end{table}

\subsection{Analysis of Branch Predictor Attacks under STBPU}
\label{analysis:stbpuattack}

An attacker possessing knowledge of their ST ($\psi /\varphi$) voids the security provided by the STBPU because they can deterministically generate outputs with any of the remapping functions used by the STBPU. 
Before we discuss how STBPU affects attacks on BPU, we show the parameters for security analysis in Table~\ref{tab:ids-st} and list several important axioms below:

\begin{enumerate}[label=A\arabic*]
\item \label{sec:a1}Attackers do not know the numerical outputs of $R_{1..4,t,p}$.

\item \label{sec:a2}Due to~\ref{sec:a1}, all the current state of the STBPU must come from detection of mispredictions and evictions.

\item \label{sec:a3}Attacker does not have inherent knowledge or control of ST of any process.
\end{enumerate}

\subsubsection{Target Injection Attacks}
\label{analysis:targetinjection}
Recall that we encrypt the targets stored in the BTB and RSB through the following means: $E_A = \varphi_{a} \oplus \tau_A$.  With Spectre V2, the attacker supplies a malicious $\tau_A$ using branch $A$ that collides with the victim's branch $V$ causing $V$ to speculate with $\tau_A$.  With the SpectreRSB, the attacker places a malicious return address $\tau_A$ on the stack that the victim speculates with. In both cases, the target the victim will use from the STBTB or STRSB is now $\tau_V = \varphi_{a} \oplus \tau_A \oplus \varphi_{v}$.  If there is a Spectre gadget located in the victim's address space at address $G$, the attack is successful if $\tau_V = G$. Due to \ref{sec:a3}, the attacker does not have knowledge or control of $\varphi_{a}$ or $\varphi_{v}$; consequently, the only variable the attacker can change is the address of $\tau_A$ to make $\tau_V = G$.  The probability that $\tau_A$ results in $\tau_V = G$ is $\frac{1}{\Omega_{STBTB}}$ or $\frac{1}{\Omega_{STRSB}}$. 
As such, the attacker must execute $\frac{\Omega_{STBTB}}{2}$ or $\frac{\Omega_{STRSB}}{2}$ different $\tau_A$ values to have a $50\%$ chance of successfully executing their target injection attack. Each incorrect $\tau_A$ will result in the misprediction counter decrementing towards zero.

\subsubsection{Reuse-based Attacks}
\label{analysis:stbpureuse}
Address mappings are randomized so that there is only a probability that an arbitrary $A$ and $V$ will collide in the STBPU.
Even though $A$ and $V$ are mapped with $R_{1..4,t,p}$, the probability that attacker branch $A$ collides with victim branch $V$ in the STBTB/STPHT is not bound by birthday attack complexity because $V$ is a static, specific address. The probability of collision is $P  ( A \Rightarrow V ) = (\frac{1}{I})(\frac{1}{TO})$. Note, we break up the probability that $A$ and $V$ are in the same set vs. the probability that $A$ and $V$ have matching tag and offsets because tag/offset comparisons are only done if $A$ and $V$ are in the same set.  This adds uncertainty for reuse-based side channels where the attacker wishes to determine the direction of $V$ since a lack of misprediction by $A$ or $V$ could mean that $A$ and $V$ do not collide, or that $V$ was not taken.  To increase the probability that an arbitrary $A$ collides with a static $V$, the attacker can execute a set of branches $S_B=\{b_1,...,b_n\}$ where $n$ is large so that one branch in $S_B$ might collide with $V$.  The probability that one of the branches in $S$ collides with $V$ is  
$P(S_B~\Rightarrow~V)~=~\sum_{i=1}^{n}P(S_{B_i}\Rightarrow~V)$.
However, noise is added using this method because it is possible that branches in $S_B$ will collide with each other.  The probability that two branches in $S_B$ collide can be approximated with birthday attack complexity because the branches in $S_B$ are arbitrary.

In order to ensure that no branches in $S_B$ collide with any other branch in $S_B$, the attacker execute the following steps: 
i) Choose a new branch $b_{new}$ with a new address in attacker's address space.
ii) For every branch $b_i$ in $S_B$, execute $b_i$ and $b_{new}$. 
iii) If no MISP. between $b_i$ and $b_{new}$, $S_B=S_B\cup\{b_{new}\}$. In order to achieve a $50\%$ probability of collision between $A$ and a branch in $S_B$, the size of $S_B$ must be $\frac{ITO}{2}$.    
The number of MISPs $M$ and evictions $E$ generated whilst generating $S_B$ of size $n=\frac{ITO}{2}$ can be approximated as follows:
\begin{equation}
\begin{split}
\label{mispeq}
M &\approx \sum_{i=0}^{n}\sum_{j=0}^{j=i}\frac{1}{\sqrt{\frac{\pi}{2}I}}\cdot\frac{1}{\sqrt{\frac{\pi}{2}TO}} = \frac{n(n+1))}{2\sqrt{\frac{\pi}{2}I}\cdot \sqrt{\frac{\pi}{2}TO}} \\
E &\approx \frac{ITO}{2} - IW
\end{split}
\end{equation}

Note the reuse-based side channel attacks on PHT do not generate evictions.  The size of the STBTB is $IW$ which is significantly smaller than $\frac{ITO}{2}$, so entries in the BTB will constantly be evicted as the attacker grows $S_B$.

Attacks such as BranchScope~\cite{evtyushkin2018branchscope} and BlueThunder~\cite{huo2019bluethunder} are viable against processors using hybrid directional predictors largely due to the inclusion of a base directional predictor in these hybrid BPUs.  Due to the complexity of TAGE tables and Perceptron weights, it is \textit{significantly} easier to maliciously modify the base directional predictor than the complex TAGE/Perceptron structures.  Since the remapping mechanisms used in our TAGE/Perceptron structures are different than the remapping functions used for the base directional predictor, little information is gained by an attacker observing mispredictions from both the base and complex directional components.  Due to \ref{sec:a1}, an attacker will not know which TAGE bank or Perceptron weight set produced a prediction.
The thresholds for re-randomization stemming from mispredictions from the directional predictor are based on the \textit{least} complex attack on the directional predictor. 
More complex attacks will be affected by re-randomization to a greater extent.

\subsubsection{Same Address Space Attacks}
\label{analysis:sasa}

Recently discovered same address space attacks~\cite{zhang2020exploring} are classified as target injection attacks, but in this case both $A$ and $V$ are located inside the attacker's address space.  As such, encrypting the target of $A$ with $\varphi_{a}$ provides no security because $V$ will decrypt $\tau_A$ with $\varphi_{a}$.  However, due to $R_i$, there is only a probability that $A$ and $V$ will collide; this probability is the same as for reuse-based attacks.  Therefore, the number of mispredictions and evictions generated while performing a same address space attack are also approximated by Equation~(\ref{mispeq}).

\subsubsection{Eviction-based Attacks}
\label{analysis:stbpueviction}
The attacker cannot deterministically create BTB eviction sets without knowing $\psi_{a}$ since address mappings change when $\psi_{a}$ is re-randomized.
With $W_{stbtb}$ ways, detecting an eviction in an arbitrary set requires $W_{stbtb} + 1$ colliding branches (same index, different tag and/or offset).
The attacker wants to fill STBTB sets so that if $V$ is executed, it disturbs one of the attacker's primed sets.
To increase the chances that $V$ will enter a primed set, the attacker must prime as many sets as possible. 
Assuming the ideal case when the attackers does not have conflicts between their own branches, they need to cover $P*I$ sets to achieve $P$ probability of a successful attack.  For example,
the probability that $A$ enters the same set as a static $V$ is $\frac{1}{I}$, so to have a $50\%$ chance of priming the set $V$ enters, the attacker must prime $\frac{I}{2}$ sets.  Na\"ively, the probability of randomly guessing $W_{stbtb}$ branches to form a single set of branches $S_e$ that enters the same STBTB set is:
\begin{equation}
P( S_e ) = \frac{1}{I^{W_{stbtb}-1}}
\end{equation}

Since this probability is not favorable, %
the attacker could apply a fast algorithm GEM~\cite{qureshi2019skewedceaser} to construct every eviction set.
The attackers uses GEM because bottom-up strategies like PPP becomes less efficient without a partitioned randomized structure~\cite{purnal21PPP} or specific cache conditions~\cite{bodduna2020brutus,bourgeat2020casa,song2021fixcache}. We assume the ideal scenario for the attacker is when most of the branches tested follow a perfect uniformity.  In this case, given a particular branch, the probability to have $W$ branches belonging to the same set is directly related to the total number of test entries.  For instance, there is a $50\%$ probability that in a group of $\frac{IW}{2}$ branches that at least $W$ branches share the same index. 
Thus, in order to achieve $P$ attack rate, the attacker needs to test at least $PIW$ branches as the initial set since the total attack lines in $L$ in GEM.
(E.g., $L>44$ for an efficient GEM in~\cite{qureshi2019skewedceaser}).
With the original setting in GEM, the attacker sets the group size $G = W+1$ and starts to eliminate groups of branches.  Although the total branch accesses will be approximately $2.3 \cdot W\cdot L$, the total eviction number will be less as the majority of the probe during each iteration will be hit.  Since the probability that each group will produce an eviction is approximately equal to $1-1/e$.  The evictions generated by testing will be negligible as $ (W+1) \cdot 1 - \frac{1}{e} \cdot n$ since the total rounds $n$ for GEM converge on the list of conflicting lines are relatively small.  However, when first placing L branches, the attacker has to trigger the same amount of evictions.
Summarizing the procedure to construct required eviction sets above, we can now 
approximate evictions numbers generated whilst building sets for $P$ attack rate as follows:
\begin{equation}
E \approx PI \times (PIW + (W+1) \times (1 - \frac{1}{e}) \times 3) \label{eq:4}
\end{equation}

\subsubsection{Re-randomization Thresholds for Baseline Model}
\label{analysis:thresholds}
STBPU has the same parameters as the baseline Intel Skylake BPU. The BTB has 8 ways and 512 sets.  The stored entries have a compressed 8-bit tag and a 5 bit offset.  The PHT has 1 way and $2^{14}$ sets.  Using Equation~\ref{mispeq}, the number of mispredictions and evictions an attacker will trigger before a successful reuse-based side channel attack on BTB is $6.9 \times 10^8$ and $\approx 2^{21}$, respectively.  Correspondingly, for a PHT reuse-based side channel, the number of triggered mispredictions is
$\approx8.38\times10^5$.
For a BTB eviction-based side channel, the average number of triggered evictions is $\frac{I}{2}$ or $5.3 \times 10^5$ per Equation~\ref{eq:4}.  For Spectre V2 and SpectreRSB, the number of triggered mispredictions is $\approx 2^{31}$. To prevent attacks, we use the lowest misprediction and eviction thresholds as the upper bounds for re-randomization of ST when evaluating the performance of STBPU.

\subsubsection{Denial-of-Service Attack on STBPU}
\label{analysis:ddos}

While the primary goal of a typical attacker is to reveal some sensitive data via a side channel or speculative execution attack,
they can also attempt to perform a denial-of-service (DoS) attack. In this attack, the goal is to cause an abnormal slowdown of a victim
process by triggering excessive branch mispredictions. We consider two DoS attack scenarios: i) Eviction-based: attacker attempts to evict 
from BPU data associated with a branch that is critical for the victim's performance. ii) Reuse-based: attacker fills BTB with bogus data hoping 
to make the victim speculatively execute code at a wrong address causing a delay due to the recovery from incorrect speculative execution.
On high level, STBPU makes both of these attacks more challenging because they rely on branch instruction collisions which
are difficult to create in STBPU. Now we will discuss each attack in more detail.

STBPU cannot eliminate the possibility of the first attack because, in STBPU, internal data structures such as BTB remain to be shared.
However, the attack becomes more difficult to carry out with STBPU. Since the victim and attacker are guaranteed to use different STs, the attacker must default to a brute force. 
Due to unknown branch-to-BTB mappings, finding eviction sets becomes a difficult task. Since BTB is a set-associative structure, to guarantee eviction 
of a certain entry, the attacker needs to find $n$ branches mapped into the same set, where $n$ is the number of ways in BTB. Since the attacker is blind, the attacker must rely on
constantly executing a large number of branches hoping to evict the victim's entries. 

The second attack is very difficult in the case of STBPU. In order to cause a hit in BTB, the attacker's and victim's branches need to have the same index, tag, and offset
after they are remapped by STBPU mechanisms with different STs. Based on our analysis above, such an event is unlikely to happen. Moreover, because the stored address
is encrypted with the ST of a different process, the predicted address would most likely point to an invalid address. Thus, erroneous speculative execution would not happen.

%% file: evaluation.tex
%% revised.

\section{Evaluation of STBPU Design}
\label{eval:main}

\begin{figure*}{}
     \centering
     \includegraphics[width=.93\textwidth]{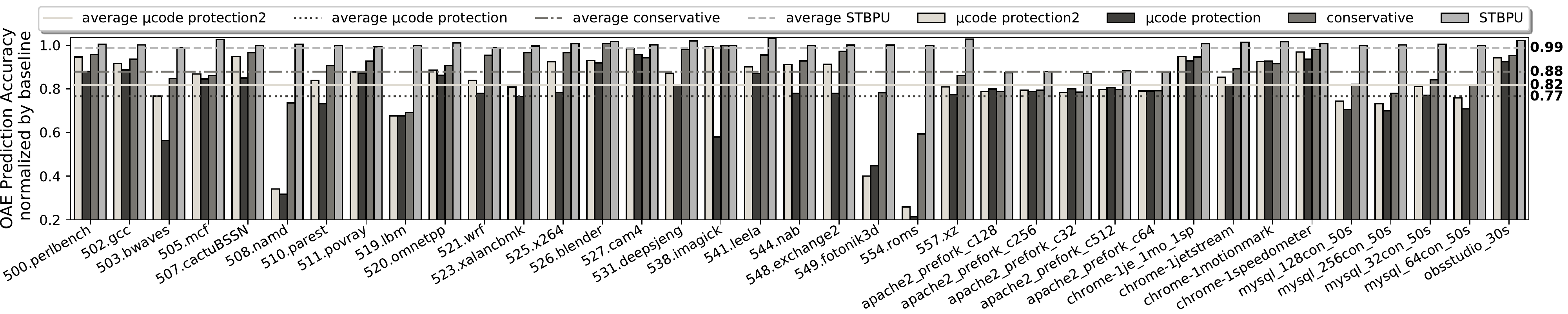}
     \caption{Overall branch prediction accuracy: STBPU against other secure BPU models}
     \label{fig:overall_accuracy_effective}
 \end{figure*}

Evaluating BPU design under realistic conditions is a challenging task. 
Sharing BPU resources creates various possibilities for branch conflicts that affect prediction accuracy. 
Moreover, some BPU protection mechanisms (e.g., Intel's IBRS) are triggered by system events such as mode and context switches.
BPU resources need to be flushed upon context/mode switches to avoid BPU training or state leakage between user and kernel processes.
workloads that involve frequent system calls and interrupts may experience performance degradation and negatively affect other
programs executing on the same core. In this situation, standard benchmark suits that are typically compute-bound and do not trigger frequent library calls, mode and context switches may not accurately evaluate BPU performance effects. Thus, a good evaluation environment needs to capture system-wide events and include real applications. A trace-based simulation is a logical choice for this. Meanwhile, the complex performance side effects caused by branch mispredictions and evictions require detailed performance data (e.g., IPC) using a cycle accurate simulator. To address both the abovementioned aspects, we evaluate STBPU using two simulation frameworks.

First, we utilize the Intel processor trace (PT) technology to collect large amounts of branch instruction traces captured from different workloads within the same CPU physical core, including user applications that cause frequent mode switches and context switches and the SPEC benchmarks.
These traces then will be passed through an in-house BPU simulator with the BPU baseline found in the Intel Skylake processor. The simulator also runs different secure models such as STBPU and reports prediction accuracy. Secondly, to evaluate fine-grained microarchitectural performance effects, we implemented the STBPU mechanisms inside gem5~\cite{gemfive} and conducted simulations in syscall-emulation (SE) mode using DerivO3CPU model 
with configurations that mimic similar modern processors.
The detailed configuration is listed in Table~\ref{tab:gem5config}. All gem5 simulations were performed simulating 110 million instructions with a warm-up of 10 million instructions.

\begin{table}[h]
\centering
\scriptsize
\begin{tabular}{|c|c|}
\hline
ISA   & Single thread: X86-64, 3.4GHz; SMT: Alpha, 3.4GHz \\ 
\hline
BPU   & BTB entries: 4096, 8-way, RAS size: 16  \\ 
\hline
Core  & 8-issue, OoO, IQ/LQ/SQ entries: 64/32/32, ROB: 192, ITLB/DTLB: 64/64  \\ 
\hline
Cache & L1-I/L1-D: 32KB/32KB both 8-way, L2: 256KB 4-way, LLC: 4MB 16-way  \\
\hline
\end{tabular}
\caption{Parameters used in gem5 simulation}
\label{tab:gem5config}
\end{table}

\subsection{Re-randomization Threshold}
\label{eval:optimization}
In Section~\ref{analysis:main}, we demonstrate the misprediction and eviction thresholds for ST re-randomization when various STBPU attacks have a $P$ attack success rate.  For BranchScope attack, to have a 50\% chance of success, the number of triggered mispredictions is estimated at 
$\approx8.38\times10^5$.  
For a BTB eviction-based side channel attack, the number of triggered evictions is $\approx$ $5.3\times10^5$.  These are the lower-bound numbers of mispredictions and evictions triggered by any attack discussed in this paper.
We aim to re-randomize ST well before the attacker has a reasonable probability of a successful attack. To do so, we utilize results from the previously discussed security analysis and derive the re-randomization thresholds as follows. We first denote the attack complexity $C$ as the least number of evictions or mispredictions that the attack needs to trigger to succeed with a 50\% chance. 
Please note that we use 50\% probability rather than 100\% since on average the attacker will succeed with half
the number of attempts needed for the fully exhaustive key search.
Let the variable $r$ be the attack difficulty factor and $\Gamma$ be the re-randomization threshold. As such, $\Gamma = r\ \cdot C$. An attack has a 50\% success rate when $r = 1$.  
For instance, if $r=0.1$, then the re-randomization thresholds for mispredictions and evictions are set to
$8.3\times10^4$ and $5.3\times10^4$, while $4.15\times10^4$ and $2.65\times10^4$ when $r=0.05$.
For further experiments, we set $r$ to 0.05 and derive the re-randomization thresholds from this value as it offers strong security guarantees with a low impact on performance.

\subsection{STBPU Performance Evaluation}
\label{eval:stbpu}

\subsubsection{Prediction Accuracy with real branch trace}
\label{eval:bpusim}
We evaluate the STBPU impact on BPU accuracy and compare it to existing na\"ive protections modeled after microcode protections based on flushing or partitioning BPU resources.  
To do so, we utilize our trace-based BPU simulator based on Intel PT technology. %
It avoids simulating the complex state of microarchitectural components. Instead, it is designed to
allow rapid testing of BPU models using branch traces from a live system running a variety of real-world scenarios. 

Each simulation instance is collected from an Intel Core i7-8550U machine that captures traces from a live physical core and includes any OS/library code executed, including naturally occurring context, mode switches, and interrupts.
This allows realistically simulating complex cross-process BPU effects and assessing how BPU flushing or ST re-randomization affects performance. 
To evaluate single-process compute-bound scenarios, we collected 23 traces from different workloads in SPEC CPU 2017. In addition, we captured traces from user and server applications, including Apache2 workloads under different prefork settings, Google Chrome traces when running single or multiple browser workloads, MySQL server, and OBS Studio.

As previously discussed, our baseline BPU model is based on recent reverse-engineering efforts~\cite{evtyushkin2016jump,kocher2018spectre,zhang2020exploring,evtyushkin2018branchscope,ret2spec,koruyeh2018spectreRSB}. To evaluate STBPU, we applied the ST mechanisms from Section~\ref{design:STBPU} to the BPU baseline model. We also created two models that mimic the baseline model with Intel's microcode-based protections, namely $\mu$code protection 1 and 2, modeling IPBP+IBRS protection with and without STIBP. Please note that microcode-based protections cannot prevent branch collisions from occurring within the same context. To prevent such collisions, more structural BPU changes are required. In particular, instead of storing compressed and truncated addresses in BTB, the full 48-bit address must be stored. As a result, the number of entries the BTB is capable of storing must be reduced (assuming unchanged hardware budget). We refer to such a model as conservative, which fully prevents any known collision-based BPU attack by flushing or partitioning. Note that STBPU achieves the same security level via customizing BPU data representations and has better performance.%

The result from simulating the above five models is demonstrated in Figure~\ref{fig:overall_accuracy_effective} 
where we aggregate all the effective predictions into a single metric: overall accuracy effective (OAE). OAE counts a branch correctly predicted if all necessary (target and direction) predictions are correct; otherwise, it's counted as mispredicted.
Figure~\ref{fig:overall_accuracy_effective} shows the overall accuracy of the various BPU models against the SPEC2017 benchmarks and user applications.
STBPU demonstrates an average 1.3\% overall effective prediction accuracy penalty.  For comparison, the microcode and the conservative BPU models suffer at least around 12\% overall accuracy loss with multiple cases of nearly 30\% reduction. With this, we conclude that based on the BPU accuracy data STBPU outperforms the microcode protections that utilize flushing and partitioning.

\subsubsection{Cycle Accurate Evaluation using gem5}

Our next evaluation focuses on the comprehensive impact of STBPU on Out-of-Order (OoO) CPU in terms of cycle accurate performance, evaluating effects of STBPU on advanced branch predictors, and SMT performance.
\begin{figure}[h]
\centering
  \begin{subfigure}{\textwidth}

 \begin{subfigure}{\textwidth}
     \includegraphics[width=.45\textwidth]{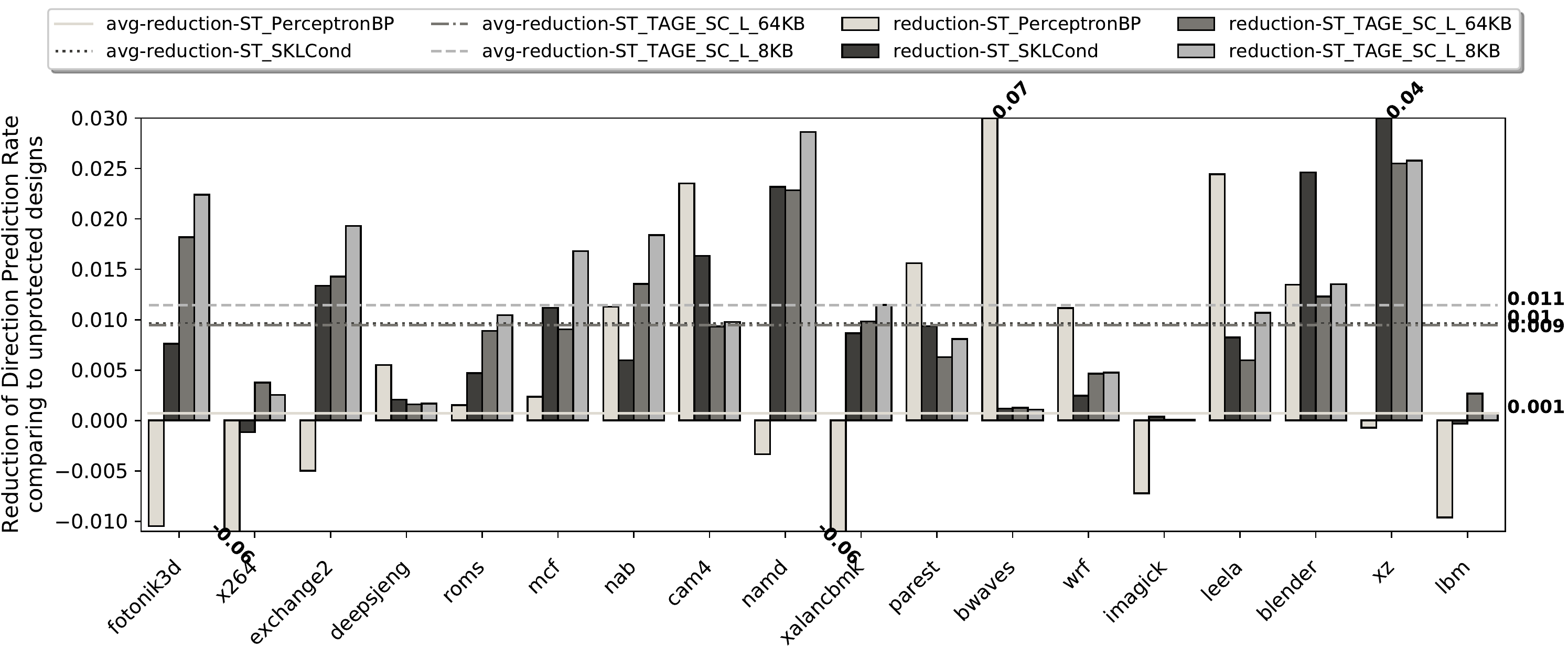}
     \label{fig:gem5_direction}
  \end{subfigure}

   \includegraphics[width=.45\textwidth]{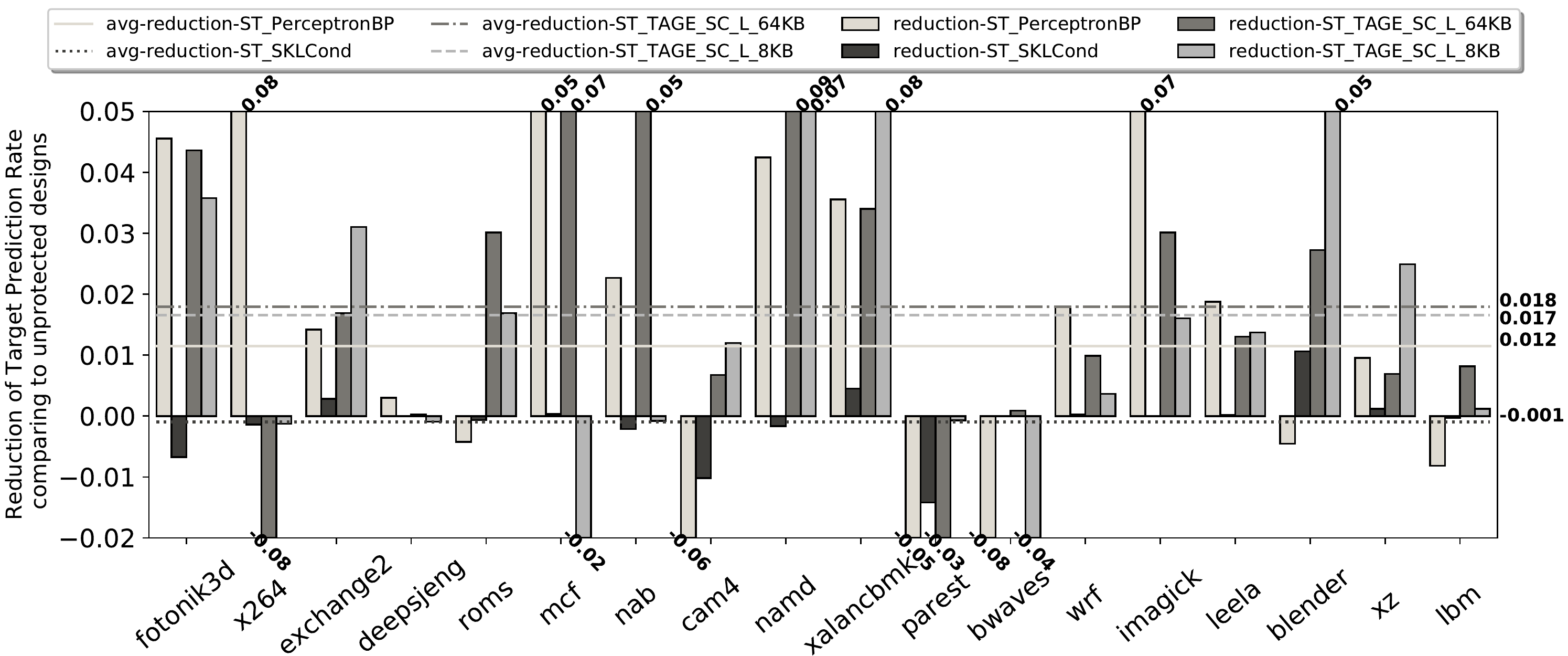}

     \label{fig:gem5_target}
  \end{subfigure}

 \begin{subfigure}{\textwidth}
     \includegraphics[width=.45\textwidth]{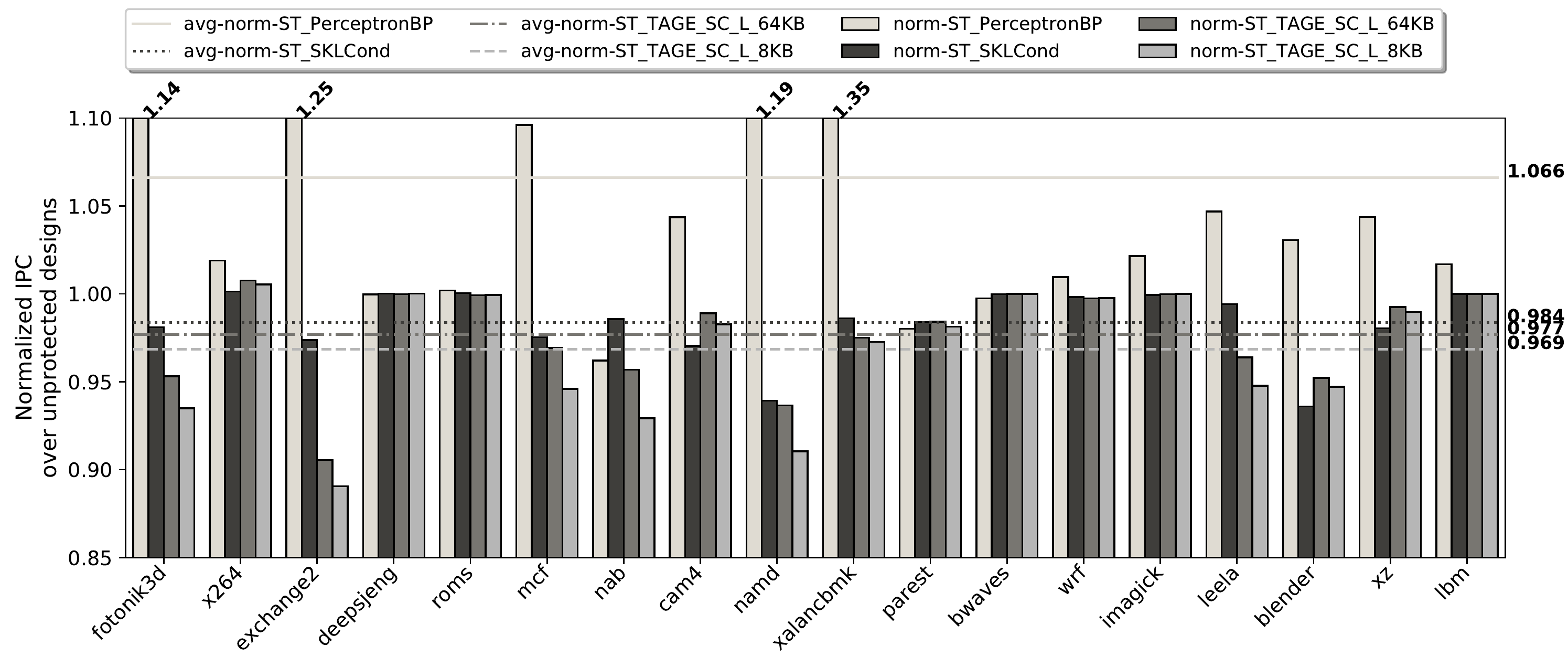}
     \label{fig:gem5_ipc}
  \end{subfigure}
  \caption{STBPU single workload evaluation in gem5}
  \label{fig:gem5_eva_nonsmt}
\end{figure}
We tested three advanced BPU models: TAGE\_SC\_L\_8KB, TAGE\_SC\_L\_64KB~\cite{tagesclAgain}, and PerceptronBP~\cite{perceptron}.  To demonstrate the consistency of accuracy between gem5 and our previous evaluation, we also ported and tested our baseline model from Section~\ref{eval:bpusim}. We refer to it as SKLCond.   
We compared the direction prediction accuracy between SKLCond in gem5 with our previous baseline model using the same workloads.  We observed on average less than 5\% direction prediction difference which validates our simulator consistency.  

We treated the aforementioned four BPUs as baseline models and implemented four STBPU models. In single process evaluations, we simulated each pair of STBPU models and their non-ST counterparts across 18 SPEC2017 workloads.  Figure~\ref{fig:gem5_eva_nonsmt} illustrates the reduction of direction / target predictions rate and the normalized IPC between STBPU designs and their non-secure counterparts.  We observe all 4 STBPU designs can achieve less than 2\% reduction on average  target prediction rate and less than 1.3\% reduction on average of direction prediction rate.  The less than 4\% average IPC reduction demonstrates the high effectiveness of STBPU designs.

\begin{figure}[t]
\centering
  \begin{subfigure}{\textwidth}

 \begin{subfigure}{\textwidth}
     \includegraphics[width=.45\textwidth]{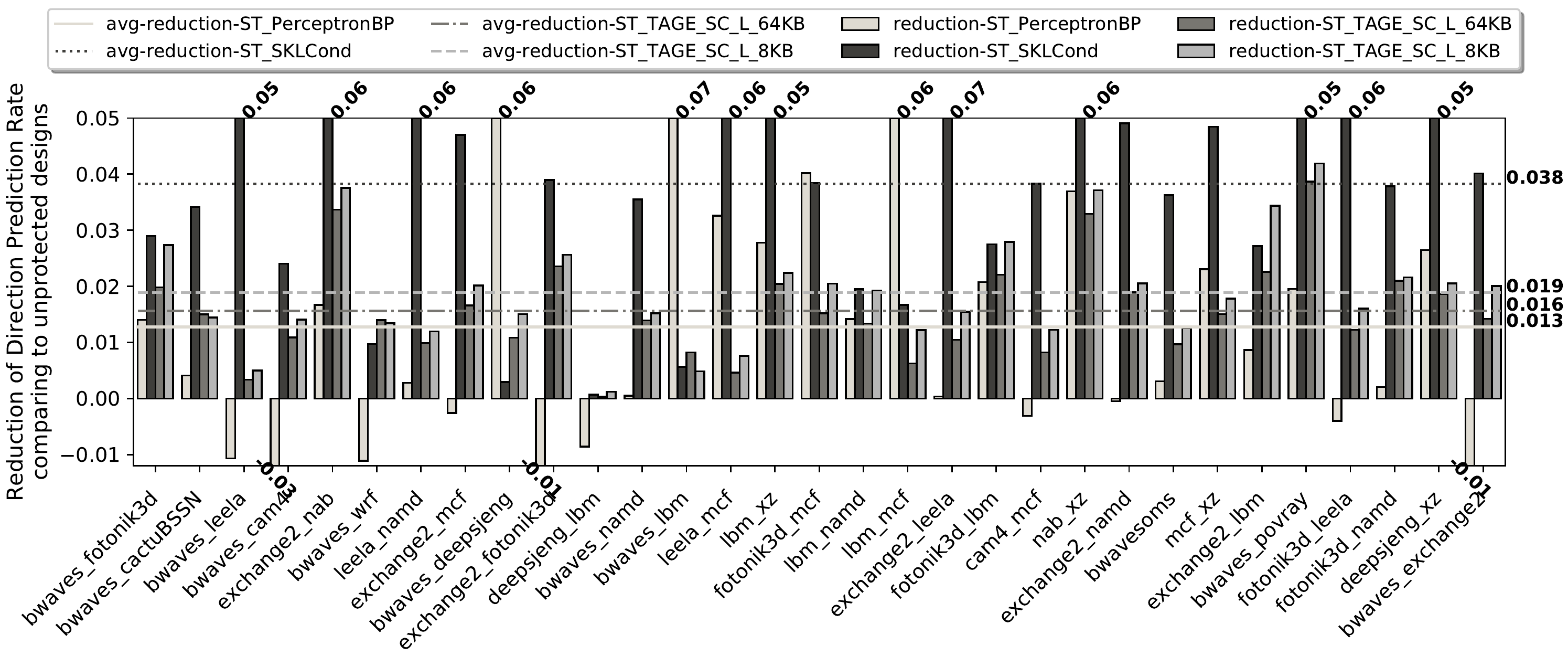}
     \label{fig:gem5_smt_direction}
  \end{subfigure}

   \includegraphics[width=.45\textwidth]{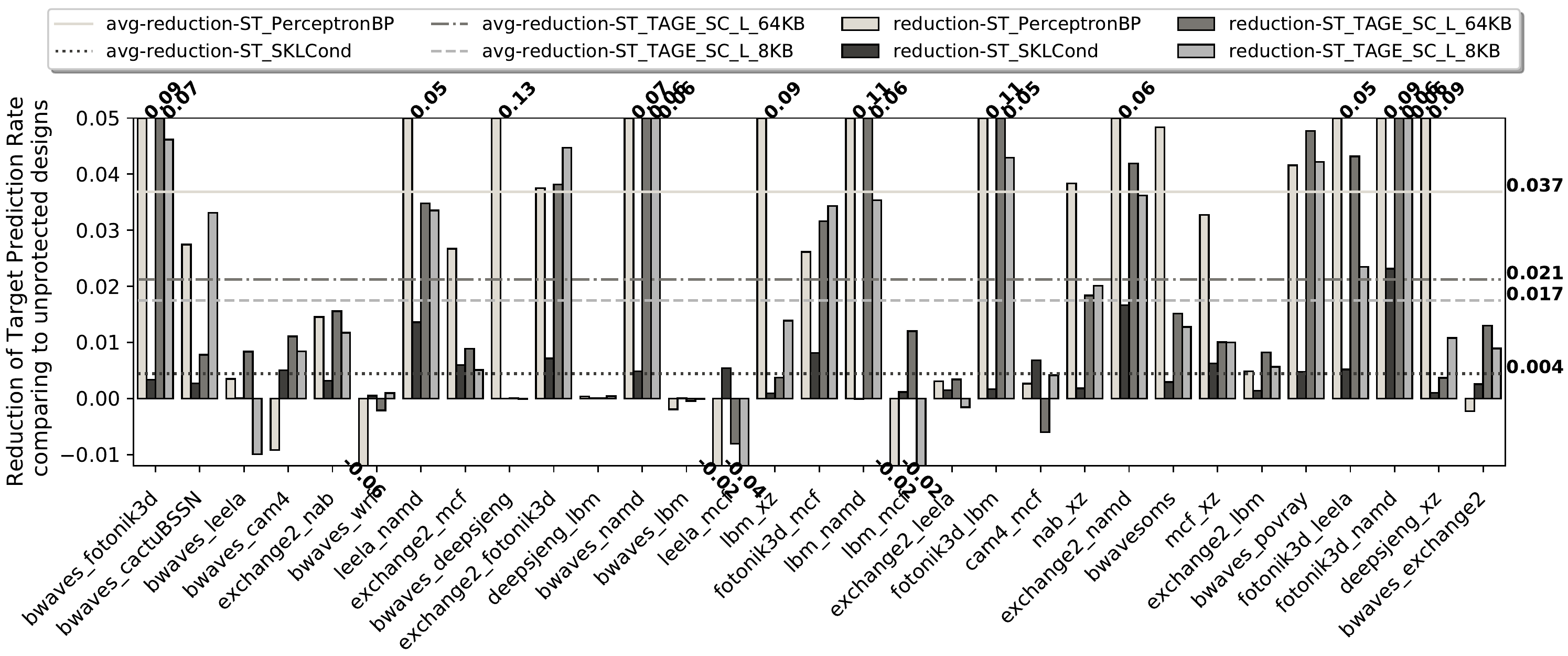}
     \label{fig:gem5_smt_target}
  \end{subfigure}

 \begin{subfigure}{\textwidth}
     \includegraphics[width=.45\textwidth]{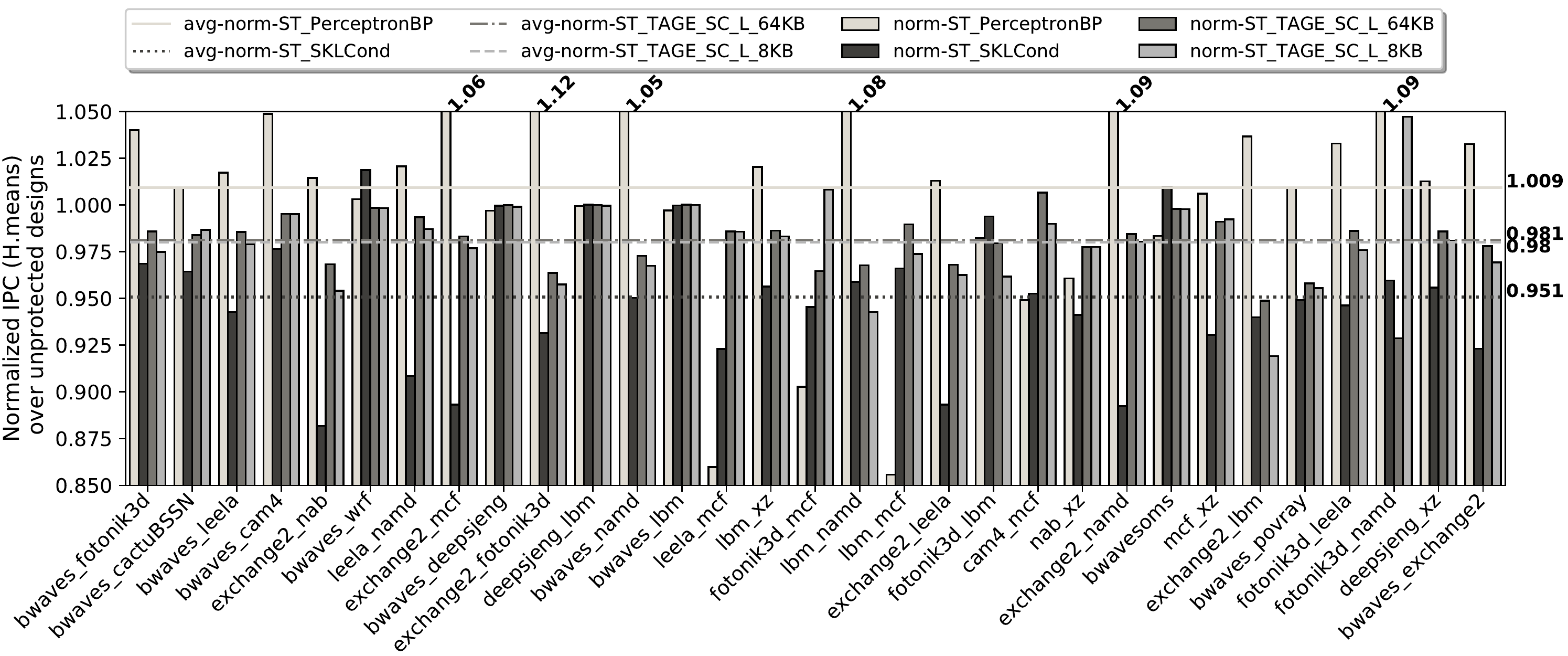}
     \label{fig:gem5_smt_ipc}
  \end{subfigure}
  \caption{SMT Evaluation of STBPU using workload pairs}

  \label{fig:gem5_eva_smt}
\end{figure}

We used the same eight BPU models in our gem5 SMT simulations.  Instead of running a single workload at a time, we grouped the individual workloads in pairs and simulated these pairs in SMT mode.  In order to accurately evaluate the STBPU impacts on overall throughput, we calculated the Harmonic means (Hmeans)~\cite{michaud2012Hmeans} of IPCs since each workload is equally valued.  Figure~\ref{fig:gem5_eva_smt} displays the overall IPC and the impact on accuracy.  We observed the ST\_SKLcond models suffer the most in SMT mode. This is because running tasks in SMT mode introduces more frequent ST re-randomizations. However, the reduction of throughput is less than 5\%.  We believe this is because the ST\_SKLcond model does not have a separate threshold register as TAGE models do for TAGE-table mispredictions. This causes more frequent direction mispredictions as shown in the first chart of Figure~\ref{fig:gem5_eva_smt}. This effect further affects the overall performance.  On the other hand, the advanced BPU models overall retain their efficiencies with minimized accuracy reduction and throughput slowdown.

\subsubsection{Aggressive ST Re-randomization and Performance}
\label{eval:future} 

It is common to see a constant arms race between protection mechanisms and more advanced attacks~\cite{purnal21PPP,qureshi2019skewedceaser}.
STBPU can withstand faster attack algorithms by reducing the ST re-randomization threshold to lower values. This would result in a more aggressive
protection scheme but can negatively affect the performance. To measure such an effect on performance, we experimented with lowering the $r$ parameter.
This is equivalent to assuming a new attack that is faster 10 times, 100 times, and even more.

\begin{figure}[]\vspace{10pt}{} %
     \centering
     \includegraphics[width=.5\textwidth]{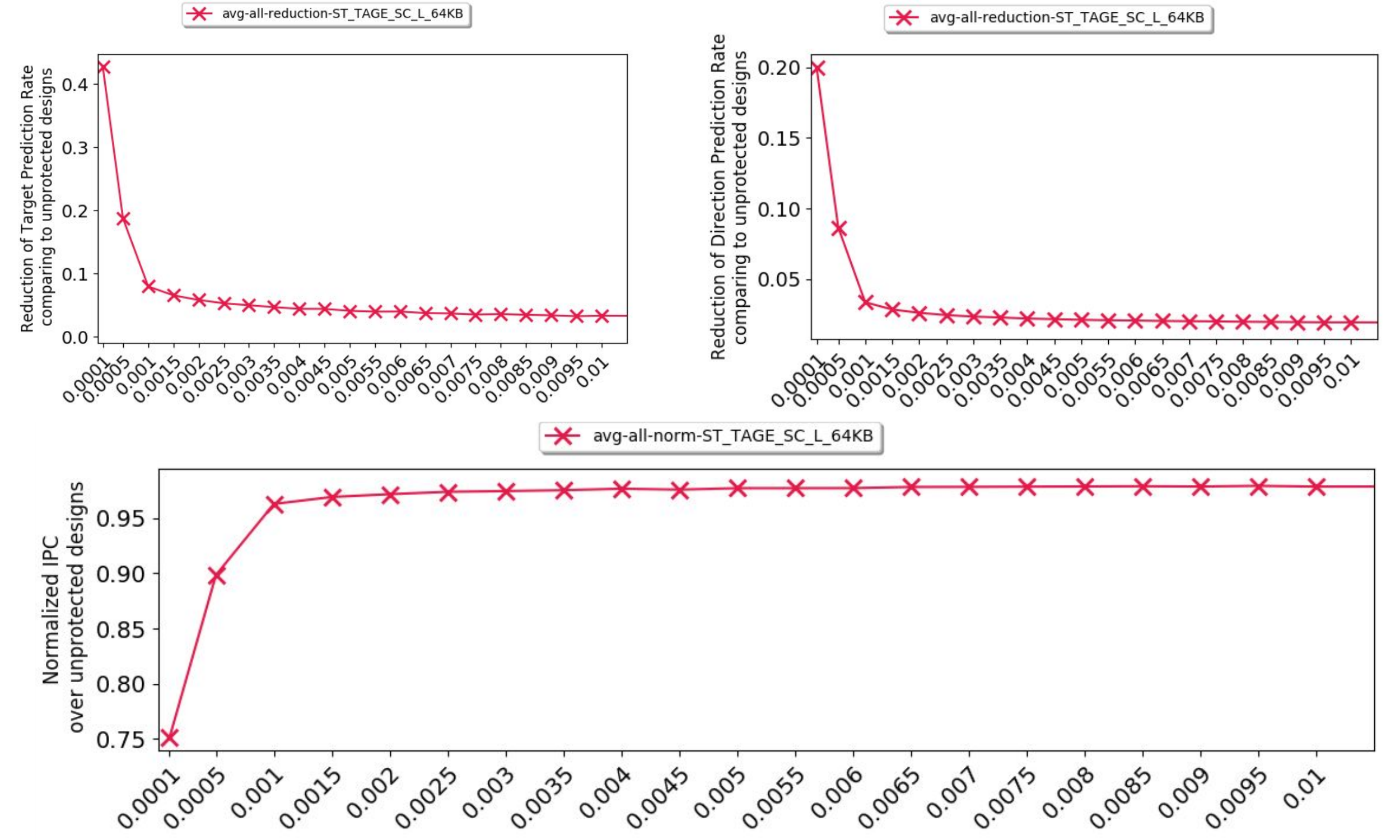}
     \caption{Effects on performance when using more aggressive re-randomization thresholds with the TAGE\_SC\_L\_64KB BPU, result are averaged from 42 combinations of SPEC CPU 2017 workload pairs. The X-axis represents the $r$ parameter.}
     \label{fig:gem5_r_smt_tagescl64}
\end{figure}

To demonstrate an extreme case, we select an advanced BPU model most sensitive to branch history loss and thus re-randomizations. We test it in the SMT setting which is more prone to trigger branch mispredictions and evictions. Figure~\ref{fig:gem5_r_smt_tagescl64} demonstrates how reducing the $r$ parameter affects the performance of the TAGE\_SC\_L BPU protected with STBPU. It shows that the thresholds can be safely reduced and maintain accuracy above 95\%. However, setting the threshold too low results in 
ST re-randomizations happening after
every few hundreds of mispredictions or evictions. This practically ceases any BPU training.

%% file: related.tex
%% seperated related work for a single section; added some latest works
\section{Related Work}
\label{related:main}

To protect against Spectre attacks, Intel processors are enhanced with Spectre-specific microcode updates, including IBRS, IBPB, and STIBP protections~\cite{intel-microcode}. The implementation of these mechanisms is not well documented and varies from one microarchitecture to another as their performance cost. Since these protections are added on top of unsafe BPU designs, they usually come with a very costly performance overhead. As a result, in practice, they are not used to their full extent and are only enabled by the OS in critical cases such as selectively protecting only a handful of processes. 

Several previous academic works proposed BPU modifications to protect against side channel and speculative execution attacks.
BRB~\cite{vougioukas2019brb} stores and reloads the entire history of the directional predictor for each process, effectively mitigating PHT collision-based attacks such as BranchScope. 
BSUP~\cite{BSUP} first encrypts the PC and then encrypts the entries of BPU, making it unsuitable for SMT processors. 

Zhao et al.~\cite{zhao2021lightweight} encode branch contents (directions and destination histories) and indexes using thread-private random numbers to achieve isolation between threads or privilege levels. Their approaches re-generate random numbers upon context and mode switches, which cannot defend against the transient execution attacks from same-address-space~\cite{zhang2020exploring,xiong2021survey}. 
Besides, our work implements ST re-randomization based
on BPU events allowing efficient branch history retention. 

The BPU in the Samsung Exynos processor is also protected with XOR-based encryption as branch history data
enhancement~\cite{grayson2020isca-exynos}. 
Since this mechanism aims to prevent speculative execution attacks such as Spectre variant 2, Exynos only encrypts stored branch targets of indirect branch instructions and returns.
However, other forms of branch collisions may still result in side channel leakage~\cite{zhao2021lightweight}. 
Additionally, in Exynos, an output of the hash function serves as a key for encrypting branch target data. It is derived from a number of process and machine-specific
inputs. 
In our work, the OS is given more flexibility for managing the ST, which allows selective branch history sharing, adjustment of re-randomization frequency, and
enforcing BPU isolation for various types of software entities such as sandboxes and libraries.

%% file: conclusion.tex
\section{Conclusion}
\label{conc:main}
We presented the STBPU, a secure branch prediction design that defends against collision-based BPU side channel and speculative execution attacks. We performed a systematization of BPU-related attacks and provided a detailed security analysis against recent attacks.  While providing security, STBPU demonstrates high performance for branch predictors modeled after real-world chips and utilizing advanced models.

%% file: acknowledgments.tex
%% added when the paper is accepted by DSN'22

\section{Acknowledgments}
The work in this paper is partially supported by Intel Corporation and National Science Foundation grant \#1850365. The statements made herein are solely the responsibility of the authors and do not necessarily reflect those of the sponsors.

%% file: main.bbl
% Generated by IEEEtranS.bst, version: 1.14 (2015/08/26)
\begin{thebibliography}{10}
\providecommand{\url}[1]{#1}
\csname url@samestyle\endcsname
\providecommand{\newblock}{\relax}
\providecommand{\bibinfo}[2]{#2}
\providecommand{\BIBentrySTDinterwordspacing}{\spaceskip=0pt\relax}
\providecommand{\BIBentryALTinterwordstretchfactor}{4}
\providecommand{\BIBentryALTinterwordspacing}{\spaceskip=\fontdimen2\font plus
\BIBentryALTinterwordstretchfactor\fontdimen3\font minus
  \fontdimen4\font\relax}
\providecommand{\BIBforeignlanguage}[2]{{%
\expandafter\ifx\csname l@#1\endcsname\relax
\typeout{** WARNING: IEEEtranS.bst: No hyphenation pattern has been}%
\typeout{** loaded for the language `#1'. Using the pattern for}%
\typeout{** the default language instead.}%
\else
\language=\csname l@#1\endcsname
\fi
#2}}
\providecommand{\BIBdecl}{\relax}
\BIBdecl

\bibitem{aciiccmez2010new}
O.~Ac{\i}i{\c{c}}mez, B.~B. Brumley, and P.~Grabher, ``{New results on
  instruction cache attacks},'' in \emph{International Workshop on
  Cryptographic Hardware and Embedded Systems}.\hskip 1em plus 0.5em minus
  0.4em\relax Springer, 2010, pp. 110--124.

\bibitem{aciiccmez2007power}
O.~Acii{\c{c}}mez, {\c{C}}.~K. Ko{\c{c}}, and J.-P. Seifert, ``{On the power of
  simple branch prediction analysis},'' in \emph{Proceedings of the 2nd ACM
  symposium on Information, computer and communications security}.\hskip 1em
  plus 0.5em minus 0.4em\relax ACM, 2007, pp. 312--320.

\bibitem{aciiccmez2007predicting}
O.~Ac{\i}i{\c{c}}mez, {\c{C}}.~K. Ko{\c{c}}, and J.-P. Seifert, ``{Predicting
  secret keys via branch prediction},'' in \emph{Cryptographers’ Track at the
  RSA Conference}.\hskip 1em plus 0.5em minus 0.4em\relax Springer, 2007, pp.
  225--242.

\bibitem{aga2017invisimem}
S.~Aga and S.~Narayanasamy, ``{InvisiMem: Smart memory defenses for memory bus
  side channel},'' in \emph{ACM SIGARCH Computer Architecture News}, vol.~45,
  no.~2.\hskip 1em plus 0.5em minus 0.4em\relax ACM, 2017, pp. 94--106.

\bibitem{barberis2022BranchHistoryInjection}
E.~Barberis, P.~Frigo, M.~Muench, H.~Bos, and C.~Giuffrida, ``{Branch History
  Injection: On the Effectiveness of Hardware Mitigations Against
  Cross-Privilege Spectre-v2 Attacks},'' in \emph{USENIX Security}, 2022.

\bibitem{bernstein2005cache}
D.~J. Bernstein, ``{Cache-timing attacks on AES},'' 2005.

\bibitem{bhattacharya2019branch}
S.~Bhattacharya, C.~Maurice, S.~Bhasin, and D.~Mukhopadhyay, ``{Branch
  Prediction Attack on Blinded Scalar Multiplication},'' \emph{IEEE
  Transactions on Computers}, vol.~69, no.~5, pp. 633--648, 2019.

\bibitem{gemfive}
N.~Binkert, B.~Beckmann, G.~Black, S.~Reinhardt, A.~Saidi, A.~Basu,
  J.~Hestness, D.~Hower, T.~Krishna, S.~Sardashti, R.~Sen, K.~Sewell, M.~Shoaib
  Bin~Altaf, N.~Vaish, M.~Hill, and D.~Wood, ``{The gem5 simulator},''
  \emph{SIGARCH Computer Architecture News}, vol.~39, pp. 1--7, 08 2011.

\bibitem{bodduna2020brutus}
R.~Bodduna, V.~Ganesan, P.~Slpsk, K.~Veezhinathan, and C.~Rebeiro, ``{Brutus:
  Refuting the security claims of the cache timing randomization countermeasure
  proposed in ceaser},'' \emph{IEEE Computer Architecture Letters}, vol.~19,
  no.~1, pp. 9--12, 2020.

\bibitem{10.1007/978-3-540-74735-2_31}
A.~Bogdanov, L.~R. Knudsen, G.~Leander, C.~Paar, A.~Poschmann, M.~J.~B.
  Robshaw, Y.~Seurin, and C.~Vikkelsoe, ``{PRESENT: An Ultra-Lightweight Block
  Cipher},'' in \emph{Cryptographic Hardware and Embedded Systems - CHES 2007},
  P.~Paillier and I.~Verbauwhede, Eds.\hskip 1em plus 0.5em minus 0.4em\relax
  Berlin, Heidelberg: Springer Berlin Heidelberg, 2007, pp. 450--466.

\bibitem{10.1007/978-3-642-23951-9_21}
A.~Bogdanov, M.~Kne{\v{z}}evi{\'{c}}, G.~Leander, D.~Toz, K.~Var{\i}c{\i}, and
  I.~Verbauwhede, ``{spongent: A Lightweight Hash Function},'' in
  \emph{Cryptographic Hardware and Embedded Systems -- CHES 2011}, B.~Preneel
  and T.~Takagi, Eds.\hskip 1em plus 0.5em minus 0.4em\relax Berlin,
  Heidelberg: Springer Berlin Heidelberg, 2011, pp. 312--325.

\bibitem{borghoff2012prince}
J.~Borghoff, A.~Canteaut, T.~G{\"u}neysu, E.~B. Kavun, M.~Knezevic, L.~R.
  Knudsen, G.~Leander, V.~Nikov, C.~Paar, C.~Rechberger \emph{et~al.},
  ``{PRINCE--a low-latency block cipher for pervasive computing
  applications},'' in \emph{International Conference on the Theory and
  Application of Cryptology and Information Security}.\hskip 1em plus 0.5em
  minus 0.4em\relax Springer, 2012, pp. 208--225.

\bibitem{bourgeat2020casa}
T.~Bourgeat, J.~Drean, Y.~Yang, L.~Tsai, J.~Emer, and M.~Yan, ``{Casa:
  End-to-end quantitative security analysis of randomly mapped caches},'' in
  \emph{2020 53rd Annual IEEE/ACM International Symposium on Microarchitecture
  (MICRO)}.\hskip 1em plus 0.5em minus 0.4em\relax IEEE, 2020, pp. 1110--1123.

\bibitem{canella2019systematic}
\BIBentryALTinterwordspacing
C.~Canella, J.~V. Bulck, M.~Schwarz, M.~Lipp, B.~von Berg, P.~Ortner,
  F.~Piessens, D.~Evtyushkin, and D.~Gruss, ``{A Systematic Evaluation of
  Transient Execution Attacks and Defenses},'' in \emph{28th {USENIX} Security
  Symposium ({USENIX} Security 19)}.\hskip 1em plus 0.5em minus 0.4em\relax
  Santa Clara, CA: {USENIX} Association, Aug. 2019, pp. 249--266. [Online].
  Available:
  \url{https://www.usenix.org/conference/usenixsecurity19/presentation/canella}
\BIBentrySTDinterwordspacing

\bibitem{chen2018sgxpectre}
G.~Chen, S.~Chen, Y.~Xiao, Y.~Zhang, Z.~Lin, and T.~H. Lai, ``{SgxPectre:
  Stealing Intel Secrets from SGX Enclaves Via Speculative Execution},'' pp.
  142--157, June 2019.

\bibitem{coppens2009practical}
B.~Coppens, I.~Verbauwhede, K.~De~Bosschere, and B.~De~Sutter, ``{Practical
  mitigations for timing-based side-channel attacks on modern x86
  processors},'' in \emph{Security and Privacy, 2009 30th IEEE Symposium
  on}.\hskip 1em plus 0.5em minus 0.4em\relax IEEE, 2009, pp. 45--60.

\bibitem{evers1996using}
M.~Evers, P.-Y. Chang, and Y.~N. Patt, ``{Using hybrid branch predictors to
  improve branch prediction accuracy in the presence of context switches},'' in
  \emph{ACM SIGARCH Computer Architecture News}, vol.~24, no.~2.\hskip 1em plus
  0.5em minus 0.4em\relax ACM, 1996, pp. 3--11.

\bibitem{evtyushkin2021computing}
D.~Evtyushkin, T.~Benjamin, J.~Elwell, J.~A. Eitel, A.~Sapello, and A.~Ghosh,
  ``{Computing with time: Microarchitectural weird machines},'' in
  \emph{Proceedings of the 26th ACM International Conference on Architectural
  Support for Programming Languages and Operating Systems}, 2021, pp. 758--772.

\bibitem{evtyushkin2016jump}
D.~Evtyushkin, D.~Ponomarev, and N.~Abu-Ghazaleh, ``{Jump over ASLR: Attacking
  branch predictors to bypass ASLR},'' in \emph{Microarchitecture (MICRO), 2016
  49th Annual IEEE/ACM International Symposium on}.\hskip 1em plus 0.5em minus
  0.4em\relax IEEE, 2016, pp. 1--13.

\bibitem{evtyushkin2016understanding}
------, ``{Understanding and mitigating covert channels through branch
  predictors},'' \emph{ACM Transactions on Architecture and Code Optimization
  (TACO)}, vol.~13, no.~1, p.~10, 2016.

\bibitem{evtyushkin2018branchscope}
D.~Evtyushkin, R.~Riley, N.~C. Abu-Ghazaleh, D.~Ponomarev \emph{et~al.},
  ``{BranchScope: A New Side-Channel Attack on Directional Branch Predictor},''
  in \emph{Proceedings of the Twenty-Third International Conference on
  Architectural Support for Programming Languages and Operating Systems}.\hskip
  1em plus 0.5em minus 0.4em\relax ACM, 2018, pp. 693--707.

\bibitem{ge2018survey}
Q.~Ge, Y.~Yarom, D.~Cock, and G.~Heiser, ``{A survey of microarchitectural
  timing attacks and countermeasures on contemporary hardware},'' \emph{Journal
  of Cryptographic Engineering}, vol.~8, no.~1, pp. 1--27, 2018.

\bibitem{grayson2020isca-exynos}
B.~Grayson, J.~Rupley, G.~Z. Zuraski, E.~Quinnell, D.~A. Jim{\'e}nez, T.~Nakra,
  P.~Kitchin, R.~Hensley, E.~Brekelbaum, V.~Sinha \emph{et~al.}, ``{Evolution
  of the samsung exynos CPU microarchitecture},'' in \emph{2020 ACM/IEEE 47th
  Annual International Symposium on Computer Architecture (ISCA)}.\hskip 1em
  plus 0.5em minus 0.4em\relax IEEE, 2020, pp. 40--51.

\bibitem{noncryptohashes}
N.~{Hua}, E.~{Norige}, S.~{Kumar}, and B.~{Lynch}, ``{Non-crypto Hardware Hash
  Functions for High Performance Networking ASICs},'' in \emph{2011 ACM/IEEE
  Seventh Symposium on Architectures for Networking and Communications
  Systems}, Oct 2011, pp. 156--166.

\bibitem{hunger2015understanding}
C.~Hunger, M.~Kazdagli, A.~Rawat, A.~Dimakis, S.~Vishwanath, and M.~Tiwari,
  ``{Understanding contention-based channels and using them for defense},'' in
  \emph{High Performance Computer Architecture (HPCA), 2015 IEEE 21st
  International Symposium on}.\hskip 1em plus 0.5em minus 0.4em\relax IEEE,
  2015, pp. 639--650.

\bibitem{huo2019bluethunder}
\BIBentryALTinterwordspacing
T.~Huo, X.~Meng, W.~Wang, C.~Hao, P.~Zhao, J.~Zhai, and M.~Li, ``{Bluethunder:
  A 2-level Directional Predictor Based Side-Channel Attack against SGX},''
  \emph{IACR Transactions on Cryptographic Hardware and Embedded Systems}, vol.
  2020, no.~1, pp. 321--347, Nov. 2019. [Online]. Available:
  \url{https://tches.iacr.org/index.php/TCHES/article/view/8401}
\BIBentrySTDinterwordspacing

\bibitem{inci2016cache}
M.~S. Inci, B.~Gulmezoglu, G.~Irazoqui, T.~Eisenbarth, and B.~Sunar, ``{Cache
  attacks enable bulk key recovery on the cloud},'' in \emph{International
  Conference on Cryptographic Hardware and Embedded Systems}.\hskip 1em plus
  0.5em minus 0.4em\relax Springer, 2016, pp. 368--388.

\bibitem{intel-microcode}
Intel, ``{Intel analysis of speculative execution side channels},'' January
  2018.

\bibitem{perceptron}
D.~A. Jim\'{e}nez and C.~Lin, ``{Dynamic Branch Prediction with Perceptrons},''
  in \emph{Proceedings of the 7th International Symposium on High-Performance
  Computer Architecture}, ser. HPCA ’01.\hskip 1em plus 0.5em minus
  0.4em\relax USA: IEEE Computer Society, 2001, p. 197.

\bibitem{kim2012stealthmem}
T.~Kim, M.~Peinado, and G.~Mainar-Ruiz, ``{STEALTHMEM: System-Level Protection
  Against Cache-Based Side Channel Attacks in the Cloud.}'' in \emph{USENIX
  Security symposium}, 2012, pp. 189--204.

\bibitem{kiriansky2018speculative}
V.~Kiriansky and C.~Waldspurger, ``{Speculative Buffer Overflows: Attacks and
  Defenses},'' \emph{arXiv preprint arXiv:1807.03757}, 2018.

\bibitem{kocher2018spectre}
P.~Kocher, D.~Genkin, D.~Gruss, W.~Haas, M.~Hamburg, M.~Lipp, S.~Mangard,
  T.~Prescher, M.~Schwarz, and Y.~Yarom, ``{Spectre attacks: Exploiting
  speculative execution},'' \emph{arXiv preprint arXiv:1801.01203}, 2018.

\bibitem{kocher1999differential}
P.~Kocher, J.~Jaffe, and B.~Jun, ``{Differential power analysis},'' in
  \emph{Annual International Cryptology Conference}.\hskip 1em plus 0.5em minus
  0.4em\relax Springer, 1999, pp. 388--397.

\bibitem{koruyeh2018spectreRSB}
E.~M. Koruyeh, K.~N. Khasawneh, C.~Song, and N.~Abu-Ghazaleh, ``{Spectre
  returns! speculation attacks using the return stack buffer},'' in \emph{12th
  USENIX Workshop on Offensive Technologies (WOOT 18)}, 2018.

\bibitem{Leander:2007:CBS:1420233.1420250}
\BIBentryALTinterwordspacing
G.~Leander and A.~Poschmann, ``{On the Classification of 4 Bit S-Boxes},'' in
  \emph{Proceedings of the 1st International Workshop on Arithmetic of Finite
  Fields}, ser. WAIFI '07.\hskip 1em plus 0.5em minus 0.4em\relax Berlin,
  Heidelberg: Springer-Verlag, 2007, pp. 159--176. [Online]. Available:
  \url{https://doi.org/10.1007/978-3-540-73074-3_13}
\BIBentrySTDinterwordspacing

\bibitem{BSUP}
\BIBentryALTinterwordspacing
J.~Lee, Y.~Ishii, and D.~Sunwoo, ``{Securing Branch Predictors with Two-Level
  Encryption},'' vol.~17, no.~3, Aug. 2020. [Online]. Available:
  \url{https://doi.org/10.1145/3404189}
\BIBentrySTDinterwordspacing

\bibitem{lee2017inferring}
S.~Lee, M.-W. Shih, P.~Gera, T.~Kim, H.~Kim, and M.~Peinado, ``{Inferring
  fine-grained control flow inside SGX enclaves with branch shadowing},'' in
  \emph{26th USENIX Security Symposium, USENIX Security}, 2017, pp. 16--18.

\bibitem{lehman2016poisonivy}
T.~S. Lehman, A.~D. Hilton, and B.~C. Lee, ``{PoisonIvy: Safe speculation for
  secure memory},'' in \emph{The 49th Annual IEEE/ACM International Symposium
  on Microarchitecture}.\hskip 1em plus 0.5em minus 0.4em\relax IEEE Press,
  2016, p.~38.

\bibitem{lipp2018meltdown}
M.~Lipp, M.~Schwarz, D.~Gruss, T.~Prescher, W.~Haas, S.~Mangard, P.~Kocher,
  D.~Genkin, Y.~Yarom, and M.~Hamburg, ``{Meltdown},'' \emph{arXiv preprint
  arXiv:1801.01207}, 2018.

\bibitem{liu2016catalyst}
F.~Liu, Q.~Ge, Y.~Yarom, F.~Mckeen, C.~Rozas, G.~Heiser, and R.~B. Lee,
  ``{Catalyst: Defeating last-level cache side channel attacks in cloud
  computing},'' in \emph{High Performance Computer Architecture (HPCA), 2016
  IEEE International Symposium on}.\hskip 1em plus 0.5em minus 0.4em\relax
  IEEE, 2016, pp. 406--418.

\bibitem{liu2015last}
F.~Liu, Y.~Yarom, Q.~Ge, G.~Heiser, and R.~B. Lee, ``{Last-level cache
  side-channel attacks are practical},'' in \emph{Security and Privacy (SP),
  2015 IEEE Symposium on}.\hskip 1em plus 0.5em minus 0.4em\relax IEEE, 2015,
  pp. 605--622.

\bibitem{m_2019}
\BIBentryALTinterwordspacing
J.~M, ``Intel® digital random number generator (drng) software implementation
  guide,'' Oct 2019. [Online]. Available:
  \url{https://software.intel.com/en-us/articles/intel-digital-random-number-generator-drng-software-implementation-guide}
\BIBentrySTDinterwordspacing

\bibitem{ret2spec}
\BIBentryALTinterwordspacing
G.~Maisuradze and C.~Rossow, ``{ret2spec: Speculative Execution Using Return
  Stack Buffers},'' \emph{CoRR}, vol. abs/1807.10364, 2018. [Online].
  Available: \url{http://arxiv.org/abs/1807.10364}
\BIBentrySTDinterwordspacing

\bibitem{maisuradze2018speculose}
------, ``{Speculose: Analyzing the Security Implications of Speculative
  Execution in CPUs},'' \emph{arXiv preprint arXiv:1801.04084}, 2018.

\bibitem{mangard2002simple}
S.~Mangard, ``{A simple power-analysis (SPA) attack on implementations of the
  AES key expansion},'' in \emph{International Conference on Information
  Security and Cryptology}.\hskip 1em plus 0.5em minus 0.4em\relax Springer,
  2002, pp. 343--358.

\bibitem{maurice2015c5}
C.~Maurice, C.~Neumann, O.~Heen, and A.~Francillon, ``{C5: cross-cores cache
  covert channel},'' in \emph{International Conference on Detection of
  Intrusions and Malware, and Vulnerability Assessment}.\hskip 1em plus 0.5em
  minus 0.4em\relax Springer, 2015, pp. 46--64.

\bibitem{menezes2018handbook}
A.~J. Menezes, P.~C. Van~Oorschot, and S.~A. Vanstone, \emph{{Handbook of
  applied cryptography}}.\hskip 1em plus 0.5em minus 0.4em\relax CRC press,
  2018.

\bibitem{messerges2002examining}
T.~S. Messerges, E.~A. Dabbish, and R.~H. Sloan, ``{Examining smart-card
  security under the threat of power analysis attacks},'' \emph{IEEE
  transactions on computers}, vol.~51, no.~5, pp. 541--552, 2002.

\bibitem{michaud2012Hmeans}
P.~Michaud, ``{Demystifying multicore throughput metrics},'' \emph{IEEE
  Computer Architecture Letters}, vol.~12, no.~2, pp. 63--66, 2012.

\bibitem{michaud1997trading}
P.~Michaud, A.~Seznec, and R.~Uhlig, ``{Trading conflict and capacity aliasing
  in conditional branch predictors},'' in \emph{ACM SIGARCH Computer
  Architecture News}, vol.~25, no.~2.\hskip 1em plus 0.5em minus 0.4em\relax
  ACM, 1997, pp. 292--303.

\bibitem{milburn2022race}
A.~Milburn, K.~Sun, and H.~Kawakami, ``{You Cannot Always Win the Race:
  Analyzing the LFENCE/JMP Mitigation for Branch Target Injection},''
  \emph{arXiv preprint arXiv:2203.04277}, 2022.

\bibitem{naghibijouybari2016covert}
H.~Naghibijouybari and N.~Abu-Ghazaleh, ``{Covert Channels on GPGPUs},''
  \emph{Computer Architecture Letters}, 2016.

\bibitem{ors2004power}
S.~B. Ors, F.~Gurkaynak, E.~Oswald, and B.~Preneel, ``{Power-Analysis Attack on
  an ASIC AES implementation},'' in \emph{Information Technology: Coding and
  Computing, 2004. Proceedings. ITCC 2004. International Conference on},
  vol.~2.\hskip 1em plus 0.5em minus 0.4em\relax IEEE, 2004, pp. 546--552.

\bibitem{osvik2006cache}
D.~A. Osvik, A.~Shamir, and E.~Tromer, ``{Cache attacks and countermeasures:
  the case of AES},'' in \emph{Cryptographers’ Track at the RSA
  Conference}.\hskip 1em plus 0.5em minus 0.4em\relax Springer, 2006, pp.
  1--20.

\bibitem{percival2005cache}
C.~Percival, ``{Cache missing for fun and profit},'' 2005.

\bibitem{prout2018measuring}
A.~Prout, W.~Arcand, D.~Bestor, B.~Bergeron, C.~Byun, V.~Gadepally, M.~Houle,
  M.~Hubbell, M.~Jones, A.~Klein \emph{et~al.}, ``{Measuring the Impact of
  Spectre and Meltdown},'' \emph{arXiv preprint arXiv:1807.08703}, 2018.

\bibitem{purnal21PPP}
A.~Purnal, L.~Giner, D.~Gruss, and I.~Verbauwhede, ``{Systematic Analysis of
  Randomization-based Protected Cache Architectures},'' in \emph{42th IEEE
  Symposium on Security and Privacy}.

\bibitem{CEASER2018}
M.~K. {Qureshi}, ``{CEASER: Mitigating Conflict-Based Cache Attacks via
  Encrypted-Address and Remapping},'' in \emph{2018 51st Annual IEEE/ACM
  International Symposium on Microarchitecture (MICRO)}, Oct 2018, pp.
  775--787.

\bibitem{qureshi2019skewedceaser}
M.~K. Qureshi, ``{New attacks and defense for encrypted-address cache},'' in
  \emph{2019 ACM/IEEE 46th Annual International Symposium on Computer
  Architecture (ISCA)}.\hskip 1em plus 0.5em minus 0.4em\relax IEEE, 2019, pp.
  360--371.

\bibitem{BallsAndBinsRaab:1998:BBS:646975.711521}
\BIBentryALTinterwordspacing
M.~Raab and A.~Steger, ``{"Balls into Bins" - A Simple and Tight Analysis},''
  in \emph{Proceedings of the Second International Workshop on Randomization
  and Approximation Techniques in Computer Science}, ser. RANDOM '98.\hskip 1em
  plus 0.5em minus 0.4em\relax Berlin, Heidelberg: Springer-Verlag, 1998, pp.
  159--170. [Online]. Available:
  \url{http://dl.acm.org/citation.cfm?id=646975.711521}
\BIBentrySTDinterwordspacing

\bibitem{raj2009resource}
H.~Raj, R.~Nathuji, A.~Singh, and P.~England, ``{Resource management for
  isolation enhanced cloud services},'' in \emph{Proceedings of the 2009 ACM
  workshop on Cloud computing security}.\hskip 1em plus 0.5em minus 0.4em\relax
  ACM, 2009, pp. 77--84.

\bibitem{ramsay2003exploring}
M.~Ramsay, C.~Feucht, and M.~H. Lipasti, ``{Exploring efficient SMT branch
  predictor design},'' in \emph{Workshop on Complexity-Effective Design, in
  conjunction with ISCA}, vol.~26, 2003.

\bibitem{Ren21isca-Iseedeaduops}
X.~Ren, L.~Moody, M.~Taram, M.~Jordan, D.~M. Tullsen, and A.~Venkat, ``{I See
  Dead $\mu$ops: Leaking Secrets via Intel/AMD Micro-Op Caches}.''

\bibitem{sakalis2019efficient}
C.~Sakalis, S.~Kaxiras, A.~Ros, A.~Jimborean, and M.~Sj{\"a}lander,
  ``{Efficient invisible speculative execution through selective delay and
  value prediction},'' in \emph{2019 ACM/IEEE 46th Annual International
  Symposium on Computer Architecture (ISCA)}.\hskip 1em plus 0.5em minus
  0.4em\relax IEEE, 2019, pp. 723--735.

\bibitem{saltaformaggio2013busmonitor}
B.~Saltaformaggio, D.~Xu, and X.~Zhang, ``{Busmonitor: A hypervisor-based
  solution for memory bus covert channels},'' \emph{Proceedings of EuroSec},
  2013.

\bibitem{schwarz2018netspectre}
M.~Schwarz, M.~Schwarzl, M.~Lipp, J.~Masters, and D.~Gruss, ``{Netspectre: Read
  arbitrary memory over network},'' pp. 279--299, 2019.

\bibitem{tagesclAgain}
A.~Seznec, ``{TAGE-SC-L Branch Predictors Again},'' 2016.

\bibitem{simakov2018effect}
N.~A. Simakov, M.~D. Innus, M.~D. Jones, J.~P. White, S.~M. Gallo, R.~L.
  DeLeon, and T.~R. Furlani, ``{Effect of Meltdown and Spectre Patches on the
  Performance of HPC Applications},'' \emph{arXiv preprint arXiv:1801.04329},
  2018.

\bibitem{song2021fixcache}
W.~Song, B.~Li, Z.~Xue, Z.~Li, W.~Wang, and P.~Liu, ``{Randomized last-level
  caches are still vulnerable to cache side-channel attacks! But we can fix
  it},'' in \emph{2021 IEEE Symposium on Security and Privacy (SP)}.\hskip 1em
  plus 0.5em minus 0.4em\relax IEEE, 2021, pp. 955--969.

\bibitem{trippel2018meltdownprime}
C.~Trippel, D.~Lustig, and M.~Martonosi, ``{MeltdownPrime and SpectrePrime:
  Automatically-Synthesized Attacks Exploiting Invalidation-Based Coherence
  Protocols},'' \emph{arXiv preprint arXiv:1802.03802}, 2018.

\bibitem{varadarajan2014scheduler}
V.~Varadarajan, T.~Ristenpart, and M.~M. Swift, ``{Scheduler-based Defenses
  against Cross-VM Side-channels.}'' in \emph{USENIX Security Symposium}, 2014,
  pp. 687--702.

\bibitem{vougioukas2019brb}
I.~Vougioukas, N.~Nikoleris, A.~Sandberg, S.~Diestelhorst, B.~M. Al-Hashimi,
  and G.~V. Merrett, ``{BRB: Mitigating Branch Predictor Side-Channels.}'' in
  \emph{2019 IEEE International Symposium on High Performance Computer
  Architecture (HPCA)}.\hskip 1em plus 0.5em minus 0.4em\relax IEEE, 2019, pp.
  466--477.

\bibitem{Wang:2007:NCD:1250662.1250723}
\BIBentryALTinterwordspacing
Z.~Wang and R.~B. Lee, ``{New Cache Designs for Thwarting Software Cache-based
  Side Channel Attacks},'' in \emph{Proceedings of the 34th Annual
  International Symposium on Computer Architecture}, ser. ISCA '07.\hskip 1em
  plus 0.5em minus 0.4em\relax New York, NY, USA: ACM, 2007, pp. 494--505.
  [Online]. Available: \url{http://doi.acm.org/10.1145/1250662.1250723}
\BIBentrySTDinterwordspacing

\bibitem{wang2008novel}
------, ``{A novel cache architecture with enhanced performance and
  security},'' in \emph{Proceedings of the 41st annual IEEE/ACM International
  Symposium on Microarchitecture}.\hskip 1em plus 0.5em minus 0.4em\relax IEEE
  Computer Society, 2008, pp. 83--93.

\bibitem{xiong2021survey}
W.~Xiong and J.~Szefer, ``{Survey of transient execution attacks and their
  mitigations},'' \emph{ACM Computing Surveys (CSUR)}, vol.~54, no.~3, pp.
  1--36, 2021.

\bibitem{yeh1991two}
T.-Y. Yeh and Y.~N. Patt, ``{Two-level adaptive training branch prediction},''
  in \emph{Proceedings of the 24th annual international symposium on
  Microarchitecture}.\hskip 1em plus 0.5em minus 0.4em\relax ACM, 1991, pp.
  51--61.

\bibitem{Micro52Jiyong-STT}
\BIBentryALTinterwordspacing
J.~Yu, M.~Yan, A.~Khyzha, A.~Morrison, J.~Torrellas, and C.~W. Fletcher,
  ``{Speculative Taint Tracking (STT): A Comprehensive Protection for
  Speculatively Accessed Data},'' in \emph{Proceedings of the 52nd Annual
  IEEE/ACM International Symposium on Microarchitecture}, ser. MICRO '52.\hskip
  1em plus 0.5em minus 0.4em\relax New York, NY, USA: Association for Computing
  Machinery, 2019, p. 954–968. [Online]. Available:
  \url{https://doi.org/10.1145/3352460.3358274}
\BIBentrySTDinterwordspacing

\bibitem{zhang2020exploring}
\BIBentryALTinterwordspacing
T.~Zhang, K.~Koltermann, and D.~Evtyushkin, ``{Exploring Branch Predictors for
  Constructing Transient Execution Trojans},'' in \emph{Proceedings of the
  Twenty-Fifth International Conference on Architectural Support for
  Programming Languages and Operating Systems}, ser. ASPLOS ’20.\hskip 1em
  plus 0.5em minus 0.4em\relax New York, NY, USA: Association for Computing
  Machinery, 2020, p. 667–682. [Online]. Available:
  \url{https://doi.org/10.1145/3373376.3378526}
\BIBentrySTDinterwordspacing

\bibitem{Zhang2015}
\BIBentryALTinterwordspacing
W.~Zhang, Z.~Bao, D.~Lin, V.~Rijmen, B.~Yang, and I.~Verbauwhede, ``{RECTANGLE:
  a bit-slice lightweight block cipher suitable for multiple platforms},''
  \emph{Science China Information Sciences}, vol.~58, no.~12, pp. 1--15, Dec
  2015. [Online]. Available: \url{https://doi.org/10.1007/s11432-015-5459-7}
\BIBentrySTDinterwordspacing

\bibitem{newSboxes}
W.~Zhang, Z.~Bao, V.~Rijmen, and M.~Liu, ``{A New Classification of 4-bit
  Optimal S-boxes and Its Application to PRESENT, RECTANGLE and SPONGENT},'' in
  \emph{Fast Software Encryption}, G.~Leander, Ed.\hskip 1em plus 0.5em minus
  0.4em\relax Berlin, Heidelberg: Springer Berlin Heidelberg, 2015, pp.
  494--515.

\bibitem{zhao2021lightweight}
L.~Zhao, P.~Li, R.~Hou, M.~C. Huang, J.~Li, L.~Zhang, X.~Qian, and D.~Meng,
  ``{A lightweight isolation mechanism for secure branch predictors},'' in
  \emph{2021 58th ACM/IEEE Design Automation Conference (DAC)}.\hskip 1em plus
  0.5em minus 0.4em\relax IEEE, 2021, pp. 1267--1272.

\end{thebibliography}
